\documentclass[3p]{elsarticle}

\usepackage[table]{xcolor}
\usepackage{graphicx}
\usepackage{enumitem}
\usepackage{multirow}
\usepackage{booktabs}
\usepackage{calc}
\usepackage{array}
\usepackage{longtable}
\usepackage[most, breakable]{tcolorbox}
\usepackage{diagbox}
\usepackage[utf8]{inputenc}
\usepackage[pdfauthor={Federico Quin, Danny Weyns, Matthias Galster, Camila Costa Silva}]{hyperref}

\newcommand{\numberofpapers}[0]{141}
\newcommand{\numberofsnowballedpapers}[0]{4}
\newcommand{\customDiagbox}[3][]{\diagbox[linecolor=white, font=\itshape\small, #1]{#2}{#3}}
\newcommand{\occurrences}[0]{\makecellslrab{\textbf{Number of}\\\textbf{occurrences}}}

\newcommand{\makecellslrab}[2][c]{%
    \renewcommand{\arraystretch}{1}\begin{tabular}[#1]{@{}c@{}}#2\end{tabular}}
\newcommand{\makecellslrableft}[2][c]{%
    \renewcommand{\arraystretch}{1}\begin{tabular}[#1]{@{}l@{}}#2\end{tabular}}

\newcommand{\makecellslrabbold}[1]{\bfseries \makecellslrab{#1}}
\graphicspath{{figures}}

\newcolumntype{M}[1]{>{\centering\arraybackslash}m{#1}}

\definecolor{new}{RGB}{20,20,200}

\newenvironment{altrowcolortbl}[1][blue!8]
{
\rowcolors{2}{white}{#1}
\begin{table}
}
{
\end{table}
}

\begin{document}

\begin{frontmatter}
\title{A/B Testing: A Systematic Literature Review}

\author[1]{Federico Quin\corref{cor1}}%
\ead{federico.quin@kuleuven.be}

\author[1,2]{Danny Weyns}
\ead{danny.weyns@kuleuven.be}

\author[3]{Matthias Galster}
\ead{matthias.galster@canterbury.ac.nz}

\author[3]{Camila Costa Silva}
\ead{camila.costasilva@pg.canterbury.ac.nz}

\cortext[cor1]{Corresponding author}
\affiliation[1]{
	organization={Distrinet, KU Leuven},
	addressline={Celestijnenlaan 200A},
	city={Leuven},
	postcode={3000},
        country={Belgium}
}
\affiliation[2]{
	organization={Linnaeus University},
	addressline={Universitetsplatsen 1},
	city={\unexpanded{V\"axj\"o}},
	postcode={351 06},
        country={Sweden}
}
\affiliation[3]{
	organization={University of Canterbury},
	addressline={69 Creyke Road},
	city={Christchurch},
	postcode={8140},
        country={New Zealand}
}

\begin{abstract}
A/B testing, also referred to as online controlled experimentation or continuous experimentation, is a form of hypothesis testing where two variants of a piece of software are compared in the field from an end user's point of view. A/B testing is widely used in practice 
to enable data-driven decision making for software development. While a few studies have explored different facets of research on A/B testing, no comprehensive study has been conducted on the state-of-the-art in A/B testing. Such a study is crucial to provide a systematic overview of the field of A/B testing driving future research forward. To address this gap and provide an overview of the state-of-the-art in A/B testing, this paper reports the results of a systematic literature review that analyzed  \numberofpapers{} primary studies. The research questions focused on the subject of A/B testing, how A/B tests are designed and executed, what roles stakeholders have in this process, and the open challenges in the area. 
Analysis of the extracted data shows that the main targets of A/B testing are algorithms, visual elements, and workflow and processes. Single classic A/B tests are the dominating type of tests, primarily based in hypothesis tests. Stakeholders have three main roles in the design of A/B tests: concept designer, experiment architect, and setup technician. The primary types of data collected during the execution of A/B tests are product/system data, user-centric data, and spatio-temporal data. The dominating use of the test results are feature selection, feature rollout, continued feature development, and subsequent A/B test design. Stakeholders have two main roles during A/B test execution: experiment coordinator and experiment assessor. The main reported open problems are related to the enhancement of proposed approaches and their usability. From our study we derived three interesting lines for future research: strengthen the adoption of statistical methods in A/B testing, improving the process of A/B testing, and enhancing the automation of A/B testing.
\end{abstract}


\end{frontmatter}

\section{Introduction}

Iterative software development and time to market are crucial to the success of software companies. Central to this is innovation by exploring new software features or experimenting with software changes. In order to enable such innovation in practice, software companies often employ A/B testing~\cite{Kohavi2017, primary-study-50, primary-study-115, primary-study-699}.
A/B testing, also referred to as online controlled experimentation or continuous experimentation, is a form of hypothesis testing where two variants of a piece of software are evaluated in the field (ranging from variants with a slightly altered GUI layout to variants of software with new features). In particular, the merit of the two variants are analyzed using metrics such as click rates of visitors of websites, members' lifetime values (LTV) in a subscription service, and user conversions in marketing~\cite{primary-study-36, Wang2019, primary-study-47}. A/B testing is extensively used in practice, including large and popular tech companies such as Google, Meta, LinkedIn, and Microsoft ~\cite{primary-study-55, Wang2019, Li2015, primary-study-44}.

Even though A/B testing is commonly used in practice, to the best of our knowledge, no comprehensive empirically grounded study has been conducted on the state-of-the-art (i.e., state-of-research in contrast to state-of-the-practice) in A/B testing. Such a study is crucial to provide a systematic overview of the field of A/B testing to drive future research forward. Three earlier studies~\cite{Auer2021, Auer2018, Ros2018} explored a number of aspects of research on A/B testing, such as research topics, type of experiments in A/B testing, and A/B tooling and metrics. Yet, these studies do not provide a comprehensive overview of the state-of-the-art that provides deeper insights in %
the types of targets to which A/B testing is applied, the roles of stakeholders in the design of A/B tests, the execution of the tests, and the usage of the test results. These insights are key to position and understand A/B testing in the broader picture of software engineering.
To tackle this issue, we performed a systematic literature review~\cite{keele2007guidelines}. %
Our study aims to provide insights on the state of research in A/B testing as a basis to guide future research. Practitioners may also benefit from the study to identify potential improvements of A/B testing in their daily practices.

The remainder of this paper is structured as follows. Section\,\ref{sec:background_and_related_work} provides a brief introduction to A/B testing and discusses related secondary studies. In Section\,\ref{sec:methodology}, we outline the research questions and summarize the methodology we used. Section\,\ref{sec:results} then presents the results, providing an answer to each research question. In Section\,\ref{sec:discussion}, we reflect on the results of the study, report insights, outline opportunities for future research, and outline threats to validity. Finally, Section\,\ref{sec:conclusions} concludes the paper.

\section{Background and related work}\label{sec:background_and_related_work}

\subsection{Background}

\begin{figure}
    \centering
    \includegraphics[width=\linewidth]{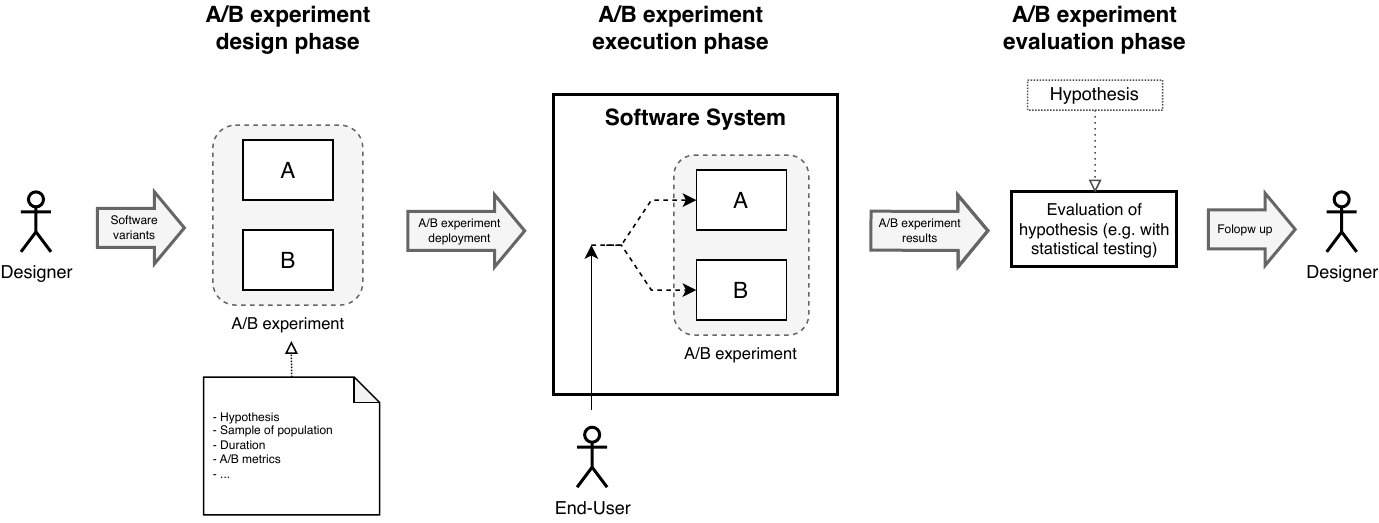}
    \caption{General A/B testing process.}
    \label{fig:ab-background}
\end{figure}

A/B testing is a method where two software variants, denoted as variant A and variant B, are compared by evaluating the merit of the variants through exposure to the end-users of the system~\cite{Siroker2013}. To compare the variants, a hypothesis is formulated together with an experiment to test it, i.e., the actual A/B test. As opposed to regular software testing, A/B testing takes place in live systems. Figure~\ref{fig:ab-background} shows the general process of A/B testing with three main phases.

The first phase of A/B testing concerns the design of an A/B test. In this experiment design, a range of parameters is specified, such as: the hypothesis, the sample of the population the experiment should be targeted to, the duration of the experiment, and the A/B metrics that are collected during the experiment. The A/B metrics are used to determine the merit of each variant during the experiment. Examples of A/B metrics include the click-through rate (CTR), number of clicks, and number of sessions~\cite{primary-study-42}.

The second phase of A/B testing consists of the execution of the A/B test in the running software system. Both variants are deployed in a live system, and the sample of the population is split among both variants. During the execution, the system keeps track of relevant data to evaluate the experiment after it finishes (according to the specified duration). Relevant data may directly correspond to the specified A/B metrics, or it may indirectly enable advanced analysis in the evaluation stage to gain additional insights from the conducted A/B tests.

The third phase of A/B testing comprises the evaluation of the experiment. After the A/B test is finished, the original hypothesis is tested, typically with a statistical test, such as a students test or Welsh's t-test~\cite{primary-study-1636, primary-study-119}. Based on the outcome of the test, the designer can then take follow-up actions, for instance initiating a rollout of a feature to the entire population or designing new A/B variants to test in subsequent A/B tests. 

\subsubsection{Controlled experiments vs A/B testing}

Traditionally, a controlled experiment is an empirical method that enables to systematically test a hypothesis~\cite{Creswell2018}. Two types of variables are distinguished in controlled experiments: independent and dependent variables. Independent variables are variables that are controlled during the experiment to test the hypothesis, for instance, a state of the art and a newly proposed approach to solve a particular design problem by a control group and a treatment group respectively. Dependent variables are variables that are measured during the experiment to compare the results of both the control and treatment group, for instance, the fault density and productivity obtained in a design task. After conducting the experiment, the hypothesis is tested and conclusions are drawn based on the results; for instance, a newly proposed design approach has a significantly lower fault density compared to the state of the art approach, but more research is required concerning the productivity. Controlled experimentation is widely used across all types of scientific fields, such as psychology~\cite{Costa2011}, pharmaceutics~\cite{Taisei2022}, education~\cite{Creswell2018}, and nowadays also in software engineering~\cite{Siegmund2015, Daneva2014, Galster2016}.

Whereas controlled experiments are typically performed \emph{offline in a controlled setting}, A/B testing uses controlled experiments to evaluate software features or variants on the \emph{end-users of a running system}. For this reason, A/B testing is often referred to as online controlled experimentation~\cite{Kohavi2020, Fabijan2020}. 
The aim of A/B testing lies in testing hypotheses in live software systems where end-users of these systems form the participants or population of the experiment. Examples of hypotheses that are tested in A/B testing often relate to improving user experience (UX)~\cite{Renz2016}, improving user interface (UI) design~\cite{Walper2020}, improving user click rates~\cite{Aharon2019}, or evaluating non functional requirements in distributed services~\cite{primary-study-327}.

\subsubsection{DevOps and A/B testing}

Development Operations (DevOps in short) has gained popularity in recent years~\cite{Feijter2017TowardsTA}. DevOps consists of a set of practices, tools, and guidelines to efficiently and effectively manage and carry out different tasks during software life-cycles. This ranges from the process of software development to the deployment and management of software at runtime. Automation of software processes plays a central part of DevOps to make life easier for developers and ease the burden of software development in general.

Common practices that are part of the DevOps lexicon are continuous integration and continuous deployment (CICD in short)~\cite{Humble2010}. CICD consists of the automation of software testing, software integration and building, and deployment of software, effectively reducing manual labor required by developers and easing the burden of deploying software. In a similar vein, continuous experimentation~\cite{Yaman2017} aims at continuously setting up experiments in software systems to test new software variants. Put differently, continuous experimentation enriches the software development process by enabling a data-driven development approach (e.g., by measuring user satisfaction of new software features early on in development). To achieve this, A/B testing is used to setup and evaluate online controlled experiments in the software system. Fabijan et al.~\cite{primary-study-50} for example perform a case study on the evolution of scaling up continuous experimentation at Microsoft, providing guidelines for other companies to conduct continuous experimentation.

\subsection{Related secondary studies}

We start with a summary of secondary studies related to the study presented in this paper. Then we pinpoint the aim of the study presented in this paper to provide a systematic overview of the state-of-the-art in A/B testing. 

\subsubsection{Summary of related reviews.}

We grouped related studies into three classes: studies with a focus on technical aspects of A/B testing, studies focusing on social aspects of A/B testing, and studies concerned with A/B testing in specific domains.

\paragraph{Technical aspects of A/B testing}

Rodriguez et al.~\cite{Rodriguez2017} performed a systematic mapping study on continuous deployment of software intensive services and products.The authors identify continuous and rapid experimentation as one of the factors that characterize continuous deployment, and elaborate on this through the lens of the deployment of these experiments and DevOps practices associated with it.
Ros and Runeson~\cite{Ros2018} put forward a mapping study on continuous experimentation and A/B testing. The authors explore research topics, organizations that employ A/B testing, and take a deeper look at the type of experimentation that is conducted. 
Auer and Felderer~\cite{Auer2018} conducted a systematic mapping study on continuous experimentation. The authors put a focus on the research topics, contributions, and research types, collaboration between industry and academia, trends in publications, popularity in publications on A/B testing, venues, and paper citations.
Recently, Auer et al.~\cite{Auer2021} presented a systematic literature review on A/B testing and continuous experimentation, leveraging the results from previous mapping studies~\cite{Ros2018, Auer2018}. The authors apply forward snowballing on a set of papers to compose the list of primary studies for the review. They then explore the core constituents of a continuous experimentation framework, and the challenges and benefits of continuous experimentation. Closely related, Erthal et al.~\cite{Erthal2022} conducted a literature review by applying an ad-hoc search, followed by snowballing on the initial set of identified papers. The study places emphasis on defining continuous experimentation and exploring its associated processes. While the authors acknowledge A/B testing as one of the strategies for achieving continuous experimentation, this literature review does not delve into the technical aspects of A/B testing.

\paragraph{Social aspects of A/B testing}

An important social aspect of A/B testing is obtaining user feedback. A significant portion of A/B tests revolves around prioritizing and optimizing the user experience. We identified two studies that focus on this social aspect. 
Fabijan et al.~\cite{Fabijan2015} present a literature review on customer feedback and data collection techniques in the context of software research and development. The authors highlight existing techniques in the literature to obtain customer feedback and organize data collection, in which software development stages the techniques are used, and what the main challenges and limitations are for the techniques. One of the techniques outlined by the authors is A/B testing, which can serve as an valuable tool to obtain user feedback on prototypes. Fabijan et al.~\cite{Fabijan2016} discuss challenges and implications of the lack of sharing customer data within large organizations. One specific case presented by the authors underpins critical issues that manifest from not sharing qualitative customer feedback in the pre-development stage with the development stage, forcing developers to repeat the collection of user feedback or to develop products without this information.

\paragraph{A/B testing in specific domains}

Beyond A/B testing at Internet-based companies, the use of A/B testing is reported in various other domains.  
An example is the domain of embedded systems. Mattos et al.~\cite{Mattos2018} explore challenges and strategies for continuous experimentation in embedded systems, providing both industrial- and research perspectives. Another domain is Cyber-Physical Systems (CPS). Giaimo et al.~\cite{Giaimo2020} present a systematic literature review on the state-of-the-art of continuous experimentation in CPS, concluding that the literature focuses more on presented challenges rather than proposing solutions to the challenges.

\paragraph{Summary}
Existing secondary studies examined A/B testing with a focus on realizing tests, associated processes, and the types of experimentation conducted. However, these studies have a particular focus, or they lack a rigorous search process to identify relevant studies. Existing studies fall short in providing insights in the \textit{target} of A/B testing (i.e., ''what'' is the subject of testing), the \textit{roles of stakeholders} in designing and executing A/B tests, and the \textit{utilization} of A/B test results.

\subsubsection{Aim of the study.}

To tackle the limitations of existing studies, we performed an in-depth literature study. We define the aim of this study using the Goal Question Metric (GQM) approach\,\cite{Basili94}: \vspace{1pt} 

\textit{Purpose}: Study and analyze

\textit{Issue}: The design and execution of A/B testing

\textit{Object}: In software systems

\textit{Viewpoint}: From the view point of researchers.\vspace{5pt} 

\noindent Concretely, we aim to investigate the subject of A/B testing, how A/B tests are designed and executed, and what the role is of stakeholders in the different phases of A/B testing. Finally, we also aim at obtaining insights in the research problems reported in the literature.

\section{Methodology}\label{sec:methodology} 

This study uses the methodology of a systematic literature review as described in~\cite{keele2007guidelines}.
This methodology describes a rigorous process to review the literature for a topic of interest. The process ensures that the review identifies, evaluates, and interprets all relevant research papers in a reproducible manner. The literature review consists of three main phases: planning, execution, and synthesis. During planning a protocol is defined for the study~\cite{research-protocol}, which includes the motivation for the study, the research questions to be answered, sources to search for papers, the search string, inclusion- and exclusion criteria, data items to be extracted from the primary studies\footnote{We use the term ``research paper'' to refer to papers that we considered for the application of inclusion and exclusion criteria in the SLR, and the term ``primary study'' for the research papers that we selected for data extraction.}, and analysis methods to be used. During execution the search string is applied as specified in the protocol, the inclusion and exclusion criteria are applied to identify the primary studies, and all the data items are extracted from these papers. Lastly, during synthesis the extracted data is analyzed and interpret to answer the research questions, and to obtain useful insights from the study.

We conducted the systematic literature review with four researchers. Further details on the process of the literature review (e.g. the roles the researchers play in the literature review) are summarized in the following sections. A complete description with the protocol, with all collected data and the data analysis are available at the study website\,\cite{research-protocol}. 

\subsection{Research questions}

To realize the aim of this study (''Study and analyze the design and execution of A/B testing in software systems from the view point of researchers.''), we put forward four research questions:

\begin{enumerate}[label=\textbf{RQ\;\!\arabic*:}, leftmargin=!, labelwidth=\widthof{\textbf{RQ4:}}]
    \item What is the subject of A/B testing?
    
    \item How are A/B tests designed? What is the role of stakeholders in this process?
    
    \item How are A/B tests executed and evaluated in the system? What is the role of stakeholders in this process? 
    
    \item What are the reported open research problems in the field of A/B testing? 
\end{enumerate}

With RQ1, we investigate the subject of A/B testing, i.e.,  the (part of the) system to which an A/B test is applied. Examples include A/B tests on program variables, application features, software components, subsystems, the system itself, and infrastructure used by the system. We also investigate the domains in which A/B testing is used.  

With RQ2, we investigate what is defined and specified in A/B tests before they are executed in the system. We look at the metrics used, whether statistical methods are used in the experiments and if so which methods, and the tools used to conduct the experiments
We also investigate which stakeholders are involved in this process and what is their role (e.g., users of the system influencing the tests that should be deployed, or architects deciding on which population the A/B tests should be run).

With RQ3, we investigate how A/B tests are executed in the system and the results are evaluated.
More specifically, we look at the way in which data is collected for evaluation in the test, the evaluation of the A/B test itself (using the collected data and, if applicable, the result of a statistical test), and the use of the test results (e.g., decision about selection of target, input for maintenance, trigger for next test in a pipeline). We also explore the role stakeholders have during this process of A/B testing (e.g., operators deciding when to finish an experiment).

With RQ4, we identify open research problems in the field of A/B testing. The problems can be derived from descriptions of limitations of proposed approaches in the reviewed papers, open challenges, or outlines of future work on A/B testing. 

\subsection{Search query}

We first identified a list of relevant terms for A/B testing from a number of known publications~\cite{King2017Book, primary-study-21, Kohavi2009, Kohavi2017, primary-study-38, primary-study-42}. We then identified and applied a gold standard~\cite{Zhang2010} to tune the terms. For a detailed description of the relevant terms and application of the gold standard, we refer to the research protocol~\cite{research-protocol}.
Figure~\ref{fig:primary-studies} (top) displays the final search query after applying the gold standard.

\subsection{Search strategy}

\begin{figure}
    \centering
    \includegraphics{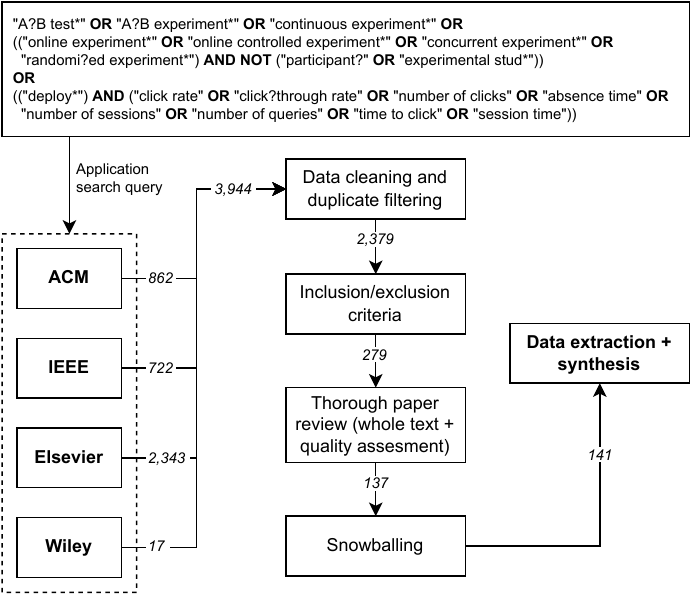}
    \caption{Primary studies selected for the systematic literature review.}
    \label{fig:primary-studies}
\end{figure}

The search query was executed October 2022. 
The search query was applied to the title and abstract of each paper in the sources (not case-sensitive). The automatic search resulted in $3,944$ papers, as shown in Figure~\ref{fig:primary-studies}. 
After filtering duplicate papers and selecting only journal versions of extensions of conference versions, $2,379$ research papers are left for further processing. 

\subsection{Search process}

After collecting the papers, we applied the following inclusion criteria:

\begin{enumerate}[label=\textbf{IC\,\!\arabic*:}, leftmargin=!, labelwidth=\widthof{\textbf{IC3:}}]
    \item Papers that either (1) have a primary focus on A/B testing (or any of its known synonyms) or (2) describe and apply (new) design(s) of A/B tests; for example introducing a proof-of-concept;
    \item Papers that include an assessment of the presented A/B tests, either by providing an evaluation through simulation with artificial data or field data, or through running one or more field experiments in a real system;
    \item Papers written in English.
\end{enumerate}

We defined IC1 such that we only include works that are relevant to the posed research questions, i.e., it is essential that the work focuses on A/B testing or their design and evaluation. Note that IC1 includes papers that address and present solutions to known challenges in A/B testing. IC2 ensured that only papers are included that contain data related to the design and/or running of A/B tests. Lastly, we only included papers that are written in English with IC3.

Besides the inclusion criteria above, we also applied the following exclusion criteria:

\begin{enumerate}[label=\textbf{EC\,\!\arabic*:}, leftmargin=!, labelwidth=\widthof{\textbf{EC4:}}]
    \item Papers that report (systematic) literature reviews, surveys (using questionnaires), interviews, and roadmap papers;
    \item Short papers ($\leq 4$ pages)\footnote{Papers published in the \textit{Lecture Notes in Computer Science} format with $< 8$ pages are also considered short.}, demos, extended abstracts, keynote talks, and tutorials;
    \item Papers with a quality score $\leq 4$ (explained in Section~\ref{subsec:data-items}).
    \item Papers that provide no or only a very brief description of the A/B testing design process or execution process.
\end{enumerate}

EC1, EC2, and EC3 excluded papers that do not directly contribute new technical advancements, preliminary works that have not been fully developed yet, or works that are not of sufficient quality. In this literature review we focus on mature, state-of-the-art research in the field of A/B testing to answer the research questions. EC4 excluded works that do not contain essential information to answer the research questions.

Papers that satisfied all inclusion criteria and none of the exclusion criteria were included as primary studies in the literature study. The application of inclusion and exclusion criteria to the titles and abstracts of the research papers resulted in $279$ papers. A thorough reading of the papers further reduced the number of papers to $137$. In addition to the research papers retrieved via the search string and filtered by applying inclusion/exclusion criteria, we applied snowballing on the cited works of these papers to capture potentially missed papers. With snowballing we discovered $\numberofsnowballedpapers{}$ additional papers, bringing the final number of primary studies to  $\numberofpapers{}$, as shown in Figure~\ref{fig:primary-studies}.

\subsection{Data items}\label{subsec:data-items}

To be able to answer the research questions, we extract the data items listed in Table~\ref{tab:data-items}. For each data item we provide a detailed description. 

\begin{table}
    \centering
    \caption{Collected data items to answer the research questions}
    \begin{tabular}{lll}
        \toprule
        \textbf{Identifier} & \textbf{Data item} & \textbf{Purpose} \\\midrule
        D1 & Authors & Documentation \\
        D2 & Year & Documentation \\
        D3 & Title & Documentation \\
        D4 & Venue & Documentation \\
        D5 & Paper type & Documentation \\
        D6 & Authors sector & Documentation \\
        D7 & Quality score & Documentation \\

        D8 & Application domain & RQ1 \\
        D9 & A/B target & RQ1 \\

        D10 & A/B test type & RQ2 \\  
        D11 & Used metrics & RQ2 \\ 
        D12 & Statistical methods employed & RQ2 \\
        D13 & Role of stakeholders in the experiment design & RQ2 \\

        D14 & Additional data collected & RQ3 \\
        D15 & Evaluation method & RQ3 \\
        D16 & Use of test results & RQ3 \\ 
        D17 & Role of stakeholder in experiment execution & RQ3 \\

        D18 & Open problems & RQ4 \\\bottomrule
    \end{tabular}
    \label{tab:data-items}
\end{table}

\begin{description}[leftmargin=!, labelwidth=\widthof{D1-D4:}]
    \item[D1-4:] Authors, year, title, and venue used for documentation purposes.
    
    \item[D5:] The type of paper. Options include: focus paper (focus on A/B testing itself, i.e., modifications, suggestions, or enhancements to the A/B testing process), or applied paper (application and evaluation of A/B testing in real software systems).
    
    \item[D6:] The sector of the authors of the primary study used for documentation (based on the author's affiliation). Options include Fully academic, Fully industrial, and Mixed.\footnote{Academic refers to affiliations that are eligible to graduate master and/or PhD students.}
    
    \item[D7:] A quality score for the reporting of the research~\cite{MahdaviHezavehi2017}. The quality score is defined on the following items: \textit{Problem definition of the study}, \textit{Problem context (relation to other work)}, \textit{Research design (study organization)}, \textit{Contributions and study results}, \textit{Derived insights}, \textit{Limitations}. Each item is rated on a scale of three levels: explicit description (2 points), general description (1 point), or no description (0 points). Therefore, the quality score is defined on a scale of 0 to 12~\cite{Madeyski2014}.
    
    \item[D8:] The application domain that is used in relation to A/B testing in the primary study. Initial options include E-commerce, Telecom, Automotive, Finance, Robotics. Further options were be derived during data collection. 
    
    \item[D9:] The target of A/B tests describes the element that is subject of A/B testing. %
    Initial options include an algorithm, a user interface, and application configurations. Further options were derived during data collection.

    \item[D10:] The type of A/B test corresponding to the number of A/B variants and the way in which they are tested. Initial options include Single (classic) A/B test, Single multivariate A/B test, Manual sequence of classic A/B tests, Manual sequence of multivariate A/B tests, Automated sequence of classic A/B tests, Automated sequence of multivariate A/B tests. Additional options were derived during data collection.
    
    \item[D11:] The metrics that are used in the A/B tests. Initial options include Click rate, Click-through rate, Number of clicks, Number of sessions, Number of queries, Absence time, Time to click, Session time. Additional options were derived during data collection.
    
    \item[D12:] The statistical method that is employed to evaluate the data obtained through the A/B test, if any. Initial options include Student test, Proportional test, No statistical test. Further options were derived during data collection. 
    
    \item[D13:] The role of stakeholders in the experiment design. Initial options include Determining A/B test goal/hypotheses, Determining A/B test duration, Tuning A/B test variants. Further options are derived during data collection.
    
    \item[D14:] Additional data that is collected during the execution of an A/B test (in addition to direct or indirect A/B metric data). Examples include User geo-location, Browser type, Timestamps of invocations or requests. Further options are derived during data collection.
    
    \item[D15:] The evaluation method used in the primary study\footnote{We distinguish data retrieved from empirical evaluation in a live system from data retrieved from simulation or an illustrative example to provide targeted insights into the execution of A/B tests during data analysis of the SLR.}. Initial options include Illustrative example, Simulation, Empirical evaluation.
    
    \item[D16:] The use of the test results gathered from A/B tests. Examples include Subsequent A/B test execution, Subsequent A/B test design, Feature rollout, Feature development. Further options are derived during data collection.
    
    \item[D17:] The role of stakeholder in the process of executing A/B tests. Initial options include A/B test alteration (adjusting individual A/B tests), A/B test triggering (starting subsequent A/B tests manually), A/B test supervision (monitoring A/B tests execution), No involvement, Unspecified. Further options are derived during data collection.
    
    \item[D18:] Reported open problems. Open problems are derived from the reported challenges, limitations, and threats to validity. Options are derived during data collection.
\end{description}

\section{Results}\label{sec:results} 

We start with the demographic information about the primary studies. Then we zoom in on each of the research questions.

\subsection{Demographic information}

Demographic information is extracted from data items Paper type (\textbf{D5}), Authors sector (\textbf{D6}), and Quality score (\textbf{D7}). 

Of the 141 primary studies, 90 (63.8\%) have a focus on A/B testing itself, while 51 (36.2\%) apply A/B testing or use it for evaluation purpose, see Table\,\ref{tab:paper_type}. 

\begin{table}
\centering
\caption{Paper types of the primary studies.}
\begin{tabular}{cc}
    \toprule
    \textbf{Type} & \occurrences{} \\\midrule
    Focus & 90 \\
    Applied & 51 \\
    \bottomrule
\end{tabular}
\label{tab:paper_type}
\end{table}

A majority of 72 primary studies (51.1\%) have industrial authors, see Table\,\ref{tab:author_background}. Forty-three studies (30.5\%) have a mix of industry and academic authors, and 26 studies (18.4\%) are from academic authors only.

\begin{table}
\centering
\caption{Author backgrounds of the primary studies.}
\begin{tabular}{cc}
    \toprule
    \textbf{Background} & \occurrences{} \\\midrule
    Academic & 26 \\
    Industry & 72 \\
    Mixed & 43 \\
    \bottomrule
\end{tabular}
\label{tab:author_background}
\end{table}

\begin{figure}
    \centering
    \includegraphics[width=.8\linewidth]{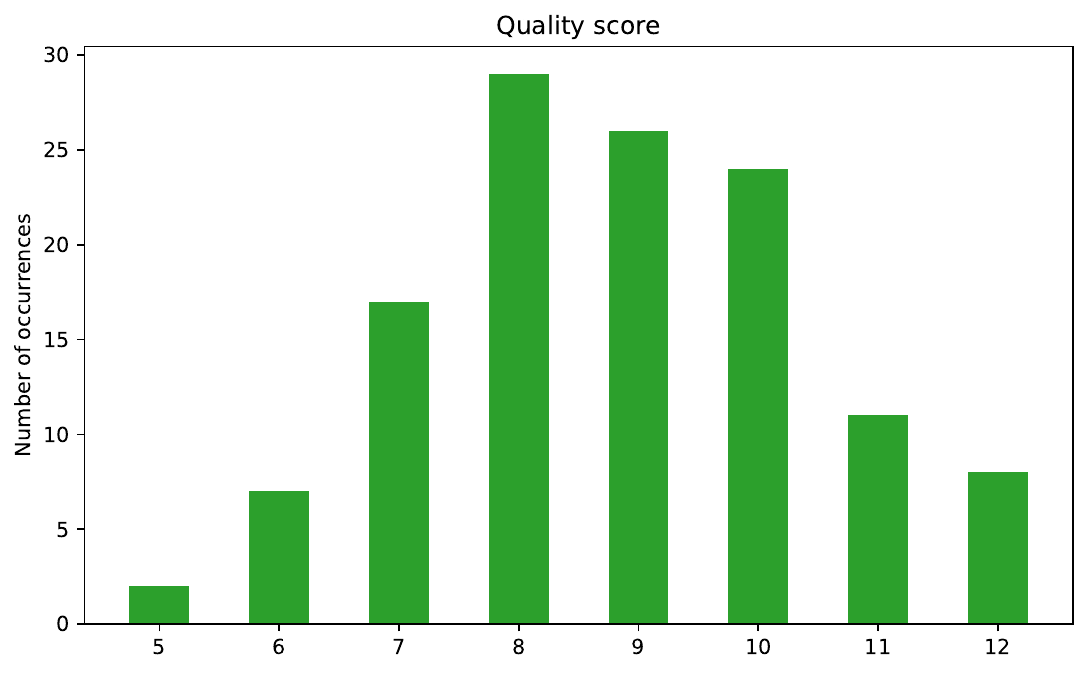}
    \caption{Quality scores of the primary studies.}
    \label{fig:quality_score}
\end{figure}

Figure\,\ref{fig:quality_score} shows the distribution of quality scores with an average of 8.81 [$\pm 1.58$]. This shows that the reporting of the research in the primary studies is of good quality. Since all papers passed the threshold of 4, none of the papers had to be excluded for the extraction of data to answer the research questions.

\subsection{RQ1: What is the subject of A/B testing?}

To answer this research question, we look at the following data items: Application domain (\textbf{D8}), and A/B target (\textbf{D9}).

\paragraph{Application domain}
 Table~\ref{tab:application_domains} lists the application domains of the primary studies.
The average number of domains is 1.13 (131 primary studies applied A/B testing in one domain, three studies in two domains, six studies in three domains, and one study in four domains). Nine studies do not mention any domain.
We observe that the most popular application domain is the Web (38 occurrences). 
Typical examples are social media platforms, such as Facebook~\cite{primary-study-77} or LinkedIn~\cite{primary-study-482}, news publishers~\cite{primary-study-105, primary-study-31}, and multimedia services, such as movie streaming at Netflix~\cite{primary-study-174}. 
The second most popular domain is search engines (35 occurrences), 
with studies conducted at Yandex~\cite{primary-study-67, primary-study-56}, Bing~\cite{primary-study-720, primary-study-100}, Yahoo~\cite{primary-study-1476, primary-study-459}, among others. 
A/B testing is also actively applied in E-commerce (27 occurrences), 
with examples from retail giant Amazon~\cite{primary-study-350}, the fashion industry~\cite{primary-study-1312}, and C2C (consumer-to-consumer) businesses, such as Etsy~\cite{primary-study-89} and Facebook marketplace~\cite{primary-study-55}.
Next we observe the application of A/B testing in what we group under ''interaction'' (22 occurrences), 
with digital communication software, such as Snap~\cite{primary-study-51} and Skype~\cite{primary-study-31}, user-operating system interaction~\cite{primary-study-293, primary-study-1796}, and application software, such as an App store~\cite{primary-study-278} and mobile games~\cite{primary-study-763}.
Lastly, we note the financial application domain (16 occurrences), 
including studies at Yahoo finance~\cite{primary-study-835} and Alipay~\cite{primary-study-1495}, transportation (4 occurrences) at for instance Didi Chuxing~\cite{primary-study-321}. Other domains are education (3 occurrences)~\cite{primary-study-227} and robotics (2 occurrences)~\cite{primary-study-1159}, among others. 

\begin{table}
\centering
\caption{Identified application domains for A/B testing.}
\begin{tabular}{lc}
    \toprule
    \makecellslrab{\textbf{Application}\\\textbf{domain}} & \occurrences{} \\\midrule	
    Web            & 38 \\
    Search engine  & 35 \\
    E-commerce     & 27 \\
    Interaction    & 22 \\
    Finances       & 16 \\
    Transportation & 4 \\
    Other          & 8 \\
    N/A            & 9 \\
    \bottomrule
\end{tabular}
\label{tab:application_domains}
\end{table}

\paragraph{A/B target} \label{sec:ab-target}

The target of the A/B test denotes the element that is subject to testing and of which (at least) two variants are compared. Table~\ref{tab:ab_targets} lists the A/B targets we identified from the primary studies, with a description and examples for each. The average number of A/B targets is 1.21 (120 primary studies applied A/B testing to one element, 26 studies to two elements, and 24 studies to three elements). Note that studies with more than one A/B target typically apply these in multiple experiments. The dominating targets of A/B testing are algorithm, visual elements, and workflow/process that together make up 86.2\% of all A/B targets reported in the primary studies. Notable, 32 primary studies did not specify a particular A/B target, for example using datasets from two prior A/B tests in the paper's evaluation without clarifying the details of these tests~\cite{primary-study-23}.

\begin{altrowcolortbl}
\centering 
\caption{Identified A/B targets, with description.}	
\renewcommand{\arraystretch}{1.5}
\begin{tabular}{m{3.3cm}m{7.5cm}c}
    \toprule
    \textbf{A/B target} & \textbf{Description} & \occurrences{} \\\midrule
    Algorithm & Updated version of an algorithm such as a recommendation algorithm~\cite{primary-study-105}, a search ranking algorithm~\cite{primary-study-52}, or an ad serving algorithm~\cite{primary-study-196}. & 58 \\
    
    Visual elements & Change to visual components such as updates to a website layout~\cite{primary-study-2042} or a general user interface update~\cite{primary-study-54}. & 33 \\
    
    Workflow / process & Alteration to the workflow of an application, e.g. the addition of a feedback button to a dashboard~\cite{primary-study-770}, or a change in a user workflow, e.g. the process of a virtual assistant tool~\cite{primary-study-2055}. & 28 \\
    
    Back-end & Optimization of a software component that is not directly visible to the user, such as testing server optimizations~\cite{primary-study-14} or adjusting application parameters for better performance~\cite{primary-study-31}. & 10 \\
    
    \makecellslrableft{New application\\functionality} & Newly introduced  functionality, such as a new widget on a web-page~\cite{primary-study-12} or additional content that is presented to the user after performing a search query~\cite{primary-study-100}. & 6 \\ 
    
    Other & This category comprises three other A/B targets: different timing and content of emails sent~\cite{primary-study-355}, varying educational resources presented to the user~\cite{primary-study-227}, and the page configuration of a web-site~\cite{primary-study-683}. & 3 \\
    
    Unspecified & The target of the A/B test was not specified in the study. & 32 \\
    \bottomrule
\end{tabular}
 \label{tab:ab_targets}
\end{altrowcolortbl}

\paragraph{Application domain vs A/B target}

We can now map the application domains with the targets of A/B testing. This analysis provides insights into which elements or components are typically the subject of A/B testing in particular domains, or alternatively which A/B targets remain unexplored in particular domains. Table~\ref{tab:application_domain_ab_target} presents this mapping. We highlight a number of key observations:

\begin{itemize}
    \item A/B testing of algorithms is applied across all application domains and for all major domains it is the primary target of A/B testing. Commonly tested algorithms include feed ranking algorithms for social media websites, recommendation algorithms for news/multimedia websites, search ranking algorithms for search engines, and advertisement serving algorithms both in the Web and search engine application domains.
    
    \item A/B testing of visual elements is particularly popular for search engines (16 studies) compared to other application domains such as Web (with only 6 studies). Typical examples include changes to font color of search engine results~\cite{primary-study-2161} and changing the position of advertisements on the result page~\cite{primary-study-209}.
    
    \item Workflow and process elements as A/B target are commonly applied across the major domains. This target is particularly popular for the Web and E-commerce (with 8 and 7 studies, respectively). Typical examples are changes to the process in which best-performing advertisements are determined in JD's advertisement platform, China's largest online retailer~\cite{primary-study-145}, and changes to the order assignment policy for on-demand meal delivery platforms~\cite{primary-study-189}. 

    \item For the Web and search engines, all types of A/B targets are applied. The main focus for the Web is on algorithms and workflow/processes, while the focus for search engines is on algorithms, visual elements, and back-end. For the Web, we notice only a single primary study with back-end as A/B target. This study targets different microservice configurations in A/B testing in order tune individual microservices for performance improvements~\cite{primary-study-727}. On the other hand, for search engines, we only noted three primary studies that target a workflow or process in A/B testing. One study evaluated a change of wording in digital advertisements~\cite{primary-study-693}, one study evaluated a change in advertisement strategies~\cite{primary-study-1636}, the last study evaluated the option to pay for ''sponsored search'' (to prioritize search results)~\cite{primary-study-785}.

    \item For e-commerce, we noticed that A/B testing is mainly used to test changes to ranking and recommendation algorithms, and to processes such as virtual assistants. Notably, we only identified a single primary study that evaluated changes to the user interface~\cite{primary-study-668}.

    \item A/B testing for back-end optimizations was identified to be most common for search engines, while we did not identify a paper in e-commerce and finances domain where A/B testing was used for back-end changes.
\end{itemize}

\begin{table}
\centering
\caption{Application domain × A/B target}
\begin{tabular}{lcccccc}
    \toprule
    \customDiagbox{Application\\domain}{A/B target} & \textbf{Algorithm} & \makecellslrab{\textbf{Visual}\\\textbf{elements}} & \makecellslrab{\textbf{Workflow}\\\textbf{/ process}} & \textbf{Back-end} & \makecellslrab{\textbf{New}\\\textbf{app.}\\\textbf{func.}} & \textbf{Other} \\\midrule 
        \textbf{Web}            & 17 & 6  & 8 & 1 & 3 & 0 \\
        \textbf{Search engine}  & 17 & 16 & 3 & 6 & 2 & 0 \\
        \textbf{E-commerce}     & 10 & 2  & 7 & 0 & 0 & 1 \\
        \textbf{Interaction}    & 5  & 6  & 2 & 2 & 1 & 0 \\
        \textbf{Finances}       & 7  & 2  & 4 & 0 & 1 & 0 \\
        \textbf{Transportation} & 2  & 0  & 0 & 1 & 1 & 0 \\
        \textbf{Other}          & 2  & 1  & 3 & 0 & 0 & 2 \\
	\bottomrule
\end{tabular}
\label{tab:application_domain_ab_target}
\end{table}

\begin{tcolorbox}[colback=white]
\textbf{Research question 1: What is the subject of A/B testing?} The main targets of A/B testing are algorithms, visual elements, workflow and processes, and back-end features. A/B testing is commonly applied in the domains of Web, search engines, e-commerce, interaction software, and finances. Algorithms are consistently tested across these domains. Visual elements are predominantly evaluated in search engines, and counter-intuitively not in e-commerce. Workflow and processes are popular A/B targets in the Web and e-commerce domains. On the other hand, back-end features such as server performance are popular targets for search engines.
\end{tcolorbox}
\subsection{RQ2: How are A/B tests designed? What is the role of stakeholders in this process?}

To answer the second research question, we look at the following data items: A/B test type (\textbf{D10}), Used metrics (\textbf{D11}), Statistical methods employed (\textbf{D12}), and Role of stakeholders in the experiment design (\textbf{D13}).

\subsubsection{Design of A/B tests}

To answer the first part of RQ2 (How are A/B tests designed?), we take a deeper look at the design of the A/B tests, focusing on the type of A/B tests, A/B metrics, and statistical methods used in the A/B tests.

\begin{figure}
    \centering
    \includegraphics[width=.8\linewidth]{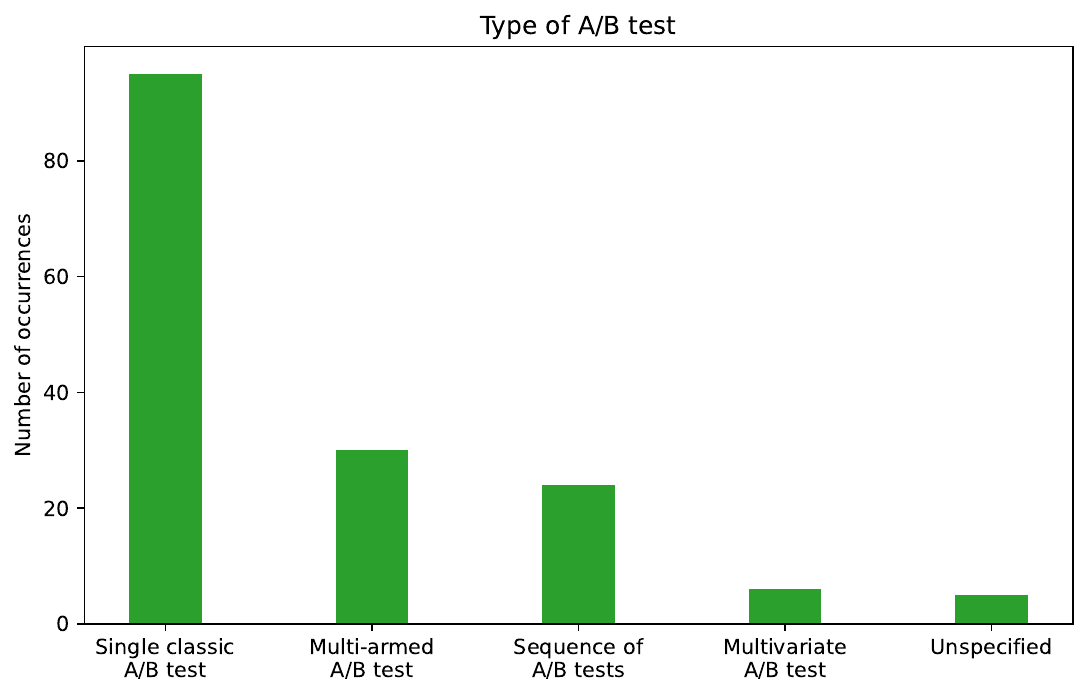}
    \caption{Identified A/B test types.}
    \label{fig:ab_test_type}
\end{figure}

\paragraph{A/B test type.} %
The type of A/B tests include single classic A/B tests with two variants, A/B test composed of more than two variants (denoted as multi-armed A/B tests), multivariate A/B test where combinations of elements are tested in one A/B test, and sequences of all the these types. Figure~\ref{fig:ab_test_type} shows the frequencies of these different A/B test types extracted from the primary studies.

Overall, we identified 155 occurrences of A/B test types, i.e., an average of 1.13 occurrences per primary study (123 studies considered a single type of A/B test, 17 studies considered two types, and one study considered three test types). 
The majority of the primary studies employed single classic A/B testing with a control variant and a treatment variant (95 occurrences). These standard test is used to test a variety of targets. 
The second most common type of A/B test is a multi-armed A/B test (30 occurrences). This type of test is composed of more than two variants under test; for example one control variant as baseline and three treatment variants with a distinct version each. These tests are commonly used to evaluate multiple versions of a recommendation algorithm, e.g.,~\cite{primary-study-760, primary-study-636}, and to test different advertisement serving algorithms, e.g.,~\cite{primary-study-436}.
The third most common type of A/B test is a sequence of classic A/B tests (24 occurrences). Examples here include the comparison of multiple variants in a manually executed sequential style (as opposed to a multi-armed A/B test where all variants are deployed simultaneously)~\cite{primary-study-362}, manually testing multiple iterations of machine learning algorithms sequentially~\cite{primary-study-60}, and automatically executing a sequence of A/B tests to handle controlled feature release in DevOps~\cite{primary-study-306}. 
The last identified A/B test type is multivariate A/B test (6 occurrences). This type of test evaluates various combinations of multiple features. As opposed to a multi-armed A/B test, a multivariate A/B test enables testing variants of more than a single feature in a singular A/B test. An example is the comparison of different combinations of varying GUI elements~\cite{primary-study-54}.

\paragraph{A/B metrics.} %
Table~\ref{tab:ab_metrics} lists the A/B metrics that we extracted from the primary studies.
In total, 493 occurrences of A/B metrics were reported in the primary studies. With a total of 198 experiments spread over 141 studies, this gives an average of 2.12 metrics per experiment\footnote{We excluded experiments and corresponding metrics of primary studies that analyzed a large number of previously conducted A/B tests.} (ranging from 1 to 8 metrics per experiment). The most common group of A/B metrics are 
engagement metrics (225 occurrences) that refer to the number of conversions\footnote{A conversion is a desired action taken in the A/B test.}, number of user sessions, time users are present on the website, and metrics related to the usage of the application or website (e.g. number of posts rated, number of bookings made).\footnote{Note that some of the primary studies do not specify explicitly the A/B metrics due to business sensitivity. Based on the available information in the study, we have included these in general engagement metrics.}
The second largest group are 
click metrics (82 occurrences). Examples include number of clicks, clicks per query, and good click rate\footnote{Good clicks are described as clicks that are meaningful during the search query session~\cite{primary-study-351}.}.
The third group of A/B metrics we identified are metrics related to monetization, i.e., revenue and cost (64 occurrences). Examples include number of purchases, order value, revenue per e-mail opening, and advertisement cost. 
%
The next group are performance metrics (50 occurrences). Examples include a simple response time of an application, bandwidth used, end-to-end latency, or playback delay of audio.
The remaining groups are metrics that track unwanted effects in the A/B tests (34 occurrences, e.g. abandonment rate or number of un-subscriptions), views (21 occurrences, e.g. number of page views or number of product views), and user feedback (17 occurrences, e.g. number of customer complaints or verbatim feedback).

\begin{table}
\centering
\caption{Identified A/B metrics.}
\begin{tabular}{lc}
    \toprule
    \textbf{A/B metric} & \occurrences{} \\\midrule
    Engagement metrics & 225 \\
    Click metrics & 82 \\
    Monetary metrics & 64 \\
    Performance metrics & 50 \\
    Negative metrics & 34 \\
    View metrics & 21 \\
    Feedback metrics & 17 \\
    \bottomrule
\end{tabular}
\label{tab:ab_metrics}
\end{table}

\paragraph{Statistical methods} %
Table~\ref{tab:statistical_methods_employed} groups the types of statistical methods used for A/B tests in the primary studies.
The most commonly used statistical method are hypothesis tests that test for equality (94 occurrences in total). The main test used in this group is a student t-test, e.g.~\cite{primary-study-1636, primary-study-699}. Other tests in this group are the Kolmogorov-Smirnov test, e.g.,~\cite{primary-study-1937}, Mann-Whitney test, e.g.,~\cite{primary-study-1679}, and Wilcoxon signed-rank test, e.g.,~\cite{primary-study-119}. Out of the 94 occurrences of this type of hypothesis test, 37 primary studies did not report the concrete test used in the analysis of the results\footnote{However, these studies did report p-values alongside the results, or explicitly refer to confidence intervals and statistically significant results of the A/B tests.}.
The second most commonly used method is bootstrapping (11 occurrences). This method constructs multiple datasets by resampling the original dataset~\cite{primary-study-67}. The newly constructed datasets are then typically used for equality hypothesis testing. The key benefit of this technique is the sensitivity improvements gained in the analysis of the results. However, a big drawback of the technique is that it is computationally expensive, especially for larger datasets~\cite{primary-study-770}.
The third mostly commonly used statistical method is a hypothesis test that tests for inference and goodness of fit (both 8 occurrences). Examples of inference hypothesis tests include using Bayesian analysis approach to ensure multiple simultaneously running experiments do not interfere~\cite{primary-study-16}, and a Bayesian approach to infer the causal effect of running ad campaigns~\cite{primary-study-198}. Examples of goodness of fit methods include sequential testing methods that are based on likelihood ratio tests~\cite{primary-study-89}, and a Wald test~\cite{primary-study-681}.
The remaining groups are correction methods (7 occurrences) with e.g. Bonferroni correction~\cite{primary-study-1587}; custom estimators for observations in A/B testing (6 occurrences), e.g., an estimator that takes variances into account~\cite{primary-study-77}; hypothesis tests for independence (5 occurrences), containing $\chi^2$ tests~\cite{primary-study-459}; and regression methods (2 occurrences), e.g. CUPED~\cite{primary-study-47}.

\begin{table}
\centering
\caption{Statistical methods employed during A/B testing.}
\begin{tabular}{lc}
    \toprule
    \textbf{Statistical methods employed} & \occurrences{} \\\midrule
    Hypothesis - equality & 57 \\
    Hypothesis - equality (concrete method unspecified) & 37 \\
    Bootstrapping & 11 \\
    Hypothesis - inference & 8 \\
    Goodness of fit & 8 \\
    Correction method & 7 \\
    Estimator & 6 \\
    Hypothesis - independence & 5 \\
    Regression method & 2 \\
    \bottomrule
\end{tabular}
\label{tab:statistical_methods_employed}
\end{table}

\subsubsection{Role of stakeholders}

To address the second part of RQ2 (What is the role of stakeholders in the design of A/B tests?), we analyze the role stakeholders play in the design of A/B tests. 

\paragraph{Roles of stakeholders} %

Table~\ref{tab:roh_design} lists the different roles of stakeholders in the design of A/B tests that we extracted from the primary studies, associated with tasks, descriptions and examples. We identified three main roles: concept designer (127 occurrences), experiment architect (111 occurrences), and setup technician (31 occurrences). The role \emph{Concept designer} consists of conceptualizing new ideas for A/B testing. The role of \emph{Experiment architect} consists of calibrating technical parameters of the experiment such as the experiment duration. The role of \emph{Setup technician} consists of taking the necessary steps required to allow the execution of the A/B test. The top task of the concept designer is designing and tuning variants of A/B tests (67 occurrences). The top task of the experiment architect is determining the duration of A/B tests (60 occurrences). Finally, the main task of the setup technician is performing post-design activities of A/B tests (25 occurrences).

\begin{altrowcolortbl}
\centering 
\caption{Roles and tasks of stakeholders in the design of A/B tests (Occ short for number of occurrences).}	
\renewcommand{\arraystretch}{1.5}
\begin{tabular}{M{.8cm}m{0.1cm}@{}m{3.5cm}m{7.3cm}c}
    \toprule
    \textbf{Role} & & \textbf{Task} & \textbf{Task description} & \textbf{Occ.} \\\midrule
    \cellcolor{white} & \cellcolor{gray!80} & Design and tune variants & Designing and tuning the variants to test. Examples are tweaking the A/B variants~\cite{primary-study-760}, or designing A/B variants for different kind of populations (e.g., old vs new users)~\cite{primary-study-679}. & 67 \\

    \cellcolor{white} & \cellcolor{gray!80} & \makecellslrableft{Determine goal or\\hypothesis} & Formulating the goal or hypothesis of the A/B test itself. Examples include the specification of a goal to find the better performing news selection algorithm~\cite{primary-study-1718} or the specification of a pre-determined hypothesis for the A/B test~\cite{primary-study-601}. & 48 \\

    \cellcolor{white} \multirow{-3}{*}[67pt]{\rotatebox[origin=c]{90}{\textbf{Concept designer (127)}}} & \cellcolor{gray!80} & \makecellslrableft{Perform pre-design\\actions} & Actions that are taken before designing the A/B test. Examples include providing motivation for A/B tests~\cite{primary-study-683} or performing offline A/B tests before moving to online A/B testing~\cite{primary-study-292}. & 12 \\

    \cellcolor{white} & \cellcolor{gray!30} & Determine duration & Determining the duration of the A/B test. Examples include choosing a fixed experiment duration (e.g., 1 week)~\cite{primary-study-601} or via an explicit expiration date~\cite{primary-study-115}. & 60 \\

    \cellcolor{white} \multirow{-2}{*}[65pt]{\rotatebox[origin=c]{90}{\textbf{Experiment architect (111)}}} & \cellcolor{gray!30} & \makecellslrableft{Determine population\\assignment} & Determining the population that should take part in the A/B test. Examples include a simple 50/50 split of all users~\cite{primary-study-476}, an assignment where the target population is determined over a two week period~\cite{primary-study-763}, or an assignment where network effects have to be taken into account~\cite{primary-study-189}. & 51 \\

    \cellcolor{white} & \cellcolor{gray!80} & \makecellslrableft{Perform post-design\\actions} & Actions that are taken after completing the design of the A/B test. Examples include performing A/A testing prior to running the A/B test~\cite{primary-study-835, primary-study-278}, validation of the A/B test design~\cite{primary-study-770}, or scheduling the execution of the A/B test~\cite{primary-study-683}. & 25 \\

    \cellcolor{white} \multirow{-2}{*}[57pt]{\rotatebox[origin=c]{90}{\textbf{Setup technician (31)}}} & \cellcolor{gray!80} & Perform metric analysis and initialization & Analyzing and potentially initializing metrics for the A/B test. An example consists of instantiating a custom A/B utility metric with negative and positive weights tied to user's actions during a search session~\cite{primary-study-100}. & 6 \\
    
    \bottomrule
\end{tabular}
\label{tab:roh_design}
\end{altrowcolortbl}

\subsubsection{Cross analysis A/B test design}

We discuss two mappings of data items: The role stakeholders take in the design of A/B tests versus A/B  test type; and the A/B metrics used in experiments versus the statistical methods employed.

\paragraph{Tasks of stakeholders vs A/B test type}

The mapping of stakeholder's tasks in the design of A/B tests across types of A/B tests is shown in~Table~\ref{tab:role_of_humans_in_the_experiment_design_ab_test_type}. 
 We observe the following:   

\begin{table}
\centering
\caption{Tasks of stakeholders × A/B test type}
\begin{tabular}{lcccc}
    \toprule
	\customDiagbox[width=\widthof{\textbf{Specify goal / hypothesis}}]{Task}{Test type (total occ.)} & \makecellslrabbold{Single classic\\A/B test (95)} & \makecellslrabbold{Multi-armed\\A/B test (30)} & \makecellslrabbold{Sequence of\\A/B tests (24)} & \makecellslrabbold{Multivariate\\A/B test (6)} \\\midrule
	\textbf{Design and tune variants} & 33 & 22 & 13 & 2 \\
	\textbf{Duration}                 & 45 & 9  & 11 & 2 \\
	\textbf{Population assignment}    & 37 & 7  & 8  & 2 \\
	\textbf{Goal/hypothesis}          & 27 & 17 & 8  & 2 \\
	\textbf{Post-design actions}      & 12 & 1  & 5  & 0 \\
	\textbf{Pre-design actions}       & 6  & 4  & 2  & 1 \\
	\textbf{Metric analysis/init.}    & 5  & 1  & 0  & 0 \\
	\bottomrule
\end{tabular}
\label{tab:role_of_humans_in_the_experiment_design_ab_test_type}
\end{table}

\begin{itemize}
    \item The primary tasks of stakeholders across all types of A/B tests are the design and tune of variants, determining the duration of experiments, the population, and the goal or hypothesis. These numbers confirm that these are essential design tasks for any A/B test.  
    \item A majority of the studies that use multi-armed A/B testing and sequence of A/B tests report the design and tuning of variants as important stakeholder task (22 and 13 occurrences respectively). Since these types of tests involve multiple variants under test, the studies often specify more details about the variants and the reasoning behind choosing which variants to test.
    \item Determining the goal or hypothesis for A/B testing is frequently mentioned for multi-armed A/B tests (17 occurrences). In contrast to conventional two-variant A/B testing that typically involves a control variant and an altered variant aimed at improving the control variant, multi-armed A/B tests involve more than two variants, so practitioners often formulate hypotheses regarding the potential performance of each variant. 
    \item Post-design actions are more often reported for sequences of A/B tests (5 instances). For instance, one primary study mentions modeling the sequence of A/B tests ~\cite{primary-study-306}, another study mentions determining the success condition of the A/B tests before executing them~\cite{primary-study-2145}, and another study refers to providing an outcome range of the A/B tests~\cite{primary-study-2275}. 
    \item Only a few primary studies report pre-design actions and metrics analysis and initialization,  independently of the type of A/B test. 
\end{itemize}

\paragraph{A/B metrics vs statistical methods used}

The statistical methods used across different types of A/B metrics are shown in  Table~\ref{tab:statistical_methods_employed_ab_netrics}.

\begin{table}
\centering
\caption{Statistical methods × A/B metrics (H short for hypothesis)}
\begin{tabular}{lccccccc}
    \toprule
    \customDiagbox[width=\widthof{\textbf{Correction method}}]{Method}{Metric} & \makecellslrabbold{Engag.} & \makecellslrabbold{Click} & \makecellslrabbold{Monetary} & \makecellslrabbold{Negative} & \makecellslrabbold{Perf.} & \makecellslrabbold{View} & \makecellslrabbold{Feedback} \\\midrule

    \textbf{H - equality} & 31 & 14 & 7 & 10 & 4 & 7 & 2 \\
    \textbf{H - equality (unsp.)} & 24 & 12 & 8 & 8 & 11 & 5 & 5 \\
    \textbf{Bootstrapping} & 9 & 2 & 2 & 3 & 3 & 1 & 1 \\
    \textbf{H - inference} & 5 & 1 & 0 & 0 & 1 & 0 & 0 \\
    \textbf{Goodness of fit} & 5 & 1 & 2 & 1 & 0 & 0 & 0 \\
    \textbf{Correction method} & 4 & 1 & 1 & 1 & 2 & 0 & 1 \\
    \textbf{Estimator} & 4 & 1 & 2 & 1 & 0 & 1 & 0 \\
    \textbf{H - independence} & 2 & 2 & 3 & 0 & 0 & 1 & 0 \\
    \textbf{Regression method} & 1 & 1 & 1 & 0 & 0 & 1 & 0 \\
	\bottomrule
\end{tabular}
\label{tab:statistical_methods_employed_ab_netrics}
\end{table}

\begin{itemize}
    \item Engagement metrics and click metrics are used across all types of statistical methods. 
    \item The concrete method used for hypothesis testing of equality is often not specified across all types of A/B metrics. 
    For monetary and performance metrics in particular, a majority of studies do not mention the concrete hypothesis testing method (8 and 11 occurrences, respectively). 
    This might be due to the sensitivity in reporting results for these types of metrics.
    \item Negative metrics are primarily used for hypothesis equality tests (10 and 8 occurrences for hypothesis equality and hypothesis equality no method specified respectively). 
    \item Hypothesis method for independence is most frequently used for the monetary metrics, yet, the use is uncommon (3 instances). 
    \item The use of feedback metrics is also uncommon and if used, the specific statistical method used is not reported (5 occurrences). 
    
\end{itemize}

\begin{tcolorbox}[colback=white]
\textbf{Research question 2: How are A/B tests designed? What is the role of stakeholders in this process?} %
The primary type of A/B test is a single classic A/B test, followed by multi-armed A/B tests and sequence of A/B test. Engagement metrics are the dominating type of A/B metrics used in A/B testing. Other prominent A/B metrics include click, monetary, and performance metrics. Hypothesis testing for equality is by far the most commonly used statistical method used in A/B testing. Remarkable, about 40\,\% of these studies that test on equality do not specify the concrete method they use for that. Stakeholders have two main roles in the design of A/B tests: concept designer and experiment architecture. Less frequently reported is a third role of setup technician.
\end{tcolorbox}

\subsection{RQ3: How are A/B tests executed? What is the role of stakeholders in this process?}

\subsubsection{Execution of A/B tests} To address the first part of RQ3 (How are A/B tests executed?), we analyze the data collected during A/B tests, the evaluation methods used, and the use of A/B tests.  

\paragraph{Data collected}

Table~\ref{tab:data_collected} lists the classes of data collected during the execution of A/B tests. We identified four types of data.
Product or system data is most commonly reported in the primary studies (48 occurrences). This data class includes the type of browser used by the end-user, the operating system of the end-user, hardware-specific information of the device used to interact with the application, and general information related to usage of the system (e.g. tracking information about item categories of products in an e-commerce application, and types of search queries processed during the A/B test).
Second most popular is user-centric data (26 occurrences). This class contains data related to how the end-user interacts with the system as well as personal information of end-users. Examples include 
scrolling characteristics of users on a web application, 
the navigation history of end-users, user feedback, and using age or current occupation of the end-user during analysis.
The third most commonly reported class is spatial-temporal data (20 occurrences) that groups data related to geographic location and time-related data. Examples include 
timestamps of requests to an application, the creation date of accounts that take part in the A/B test, and spatial information such as the country and region of end-users.
Lastly, a few primary studies report the use of secondary data (6 occurrences). Data in this class correspond to A/B metrics that do not serve as main evaluation metrics for A/B tests. Examples are 
the number of clicks or page views that are used for additional analysis after conducting the A/B tests.

\begin{table}
\centering
\caption{Data collected for the A/B tests.}
\begin{tabular}{lc}
    \toprule
    \textbf{Data collected} & \occurrences{} \\\midrule
    Product/system data & 48 \\
    User-centric data & 26 \\
    Spatial-temporal data & 20 \\
    Secondary data & 6 \\
    \bottomrule
\end{tabular}
\label{tab:data_collected}
\end{table}

\paragraph{Evaluation method}

Table~\ref{tab:evaluation_method} summarizes the identified evaluation methods.
The vast majority of primary studies provide results from an empirical evaluation (100 occurrences), i.e., executing A/B tests in live systems. A substantial number of studies use historical data from previously conducted A/B tests to simulate new A/B tests (26 occurrences), while a handful of studies (15 occurrences) use simulations without historical 
data as their evaluation method. Lastly, a few studies use illustrative examples (10 occurrences), case studies (5 occurrences), and a single primary study provides a theoretical evaluation~\cite{primary-study-209}.

\begin{table}
\centering
\caption{Evaluation method used in the primary studies.}
\begin{tabular}{lc}
    \toprule
    \textbf{Evaluation method} & \occurrences{} \\\midrule
    Empirical evaluation & 100 \\
    Simulation based on real empirical data & 26 \\
    Simulation & 15 \\
    Illustrative example & 10 \\
    Case study & 5 \\
    Theoretical & 1 \\
    \bottomrule
\end{tabular}
\label{tab:evaluation_method}
\end{table}

\paragraph{Use of test results}

Table~\ref{tab:use_of_test_results} lists the use of test results extracted from the primary studies. Use of test results refers to what stakeholders do with the obtained data and analyses of A/B tests, such using the results to design additional A/B tests. As the table shows, the main usages of A/B test results are the selection and rollout of a feature (71 and 24 occurrences respectively). A number of studies aim at validating the effectiveness of the A/B testing process itself (12 occurrences). The use of test results to trigger a subsequent A/B test seems not very well explored (4 occurrences).

\begin{altrowcolortbl}
\centering
\caption{Use of test results gathered from A/B test execution.}
\renewcommand{\arraystretch}{1.5}
\begin{tabular}{m{3.6cm}m{8.2cm}c}
    \toprule
    \textbf{Use of test results} & \textbf{Description} & \textbf{Occur.} \\\midrule 
    Feature selection & The results of the A/B test are used to determine which variant presents an improvement to the application. Examples include selecting a new version of a ranking algorithm~\cite{primary-study-2026, primary-study-12} or a recommendation algorithm~\cite{primary-study-473}, and selecting a different visual design~\cite{primary-study-1337}. & 71 \\
    
    Feature rollout & The results of the A/B test are used to determine if the rollout of a feature should be continued or halted, as for example outlined by practitioners at Microsoft~\cite{primary-study-615, primary-study-1965}.  & 24 \\
    
    \makecellslrableft{Continue feature\\ development} & The results of the A/B test are used as a driving force for further feature development, e.g. fine-tuning newly proposed A/B metrics based on periodicity patterns after obtaining promising results~\cite{primary-study-56}, and further developing personalization methods~\cite{primary-study-1476}. & 17 \\
    
    \makecellslrableft{Subsequent A/B\\test design} & The results of the A/B test are used for future A/B test design, for example suggesting alternative A/B variants to test in future A/B tests~\cite{primary-study-2055}, and designing a new A/B test to further test the quality of an A/B metric prediction model\footnote{The prediction model was trained on data from the previously conducted A/B tests.}~\cite{primary-study-561}. & 15 \\
    
    \makecellslrableft{Validation effectiveness\\ of A/B testing process} & The results of the A/B test are used to demonstrate the effectiveness of the newly proposed or improved A/B testing approach by the authors. Examples include evaluating a newly proposed counterfactual framework to run seller-side A/B tests in two-sided marketplaces~\cite{primary-study-55}, and the validation of a new statistical methodology for continuous monitoring of A/B tests~\cite{primary-study-36}. & 12 \\
    
    \makecellslrableft{Validation of a\\research question} & A/B testing is used to validate a research question put forward by the authors. One example consists of investigating the hypothesis under which circumstances companies should pay for advertising in search engines~\cite{primary-study-785}. & 10 \\
    
    Bug detection / fixing & The results of the A/B test are used to detect potential bugs or validate bug fixes, e.g. probing for data quality issues in A/B tests of ML models to uncover potential bugs~\cite{primary-study-60}. & 5 \\
    
    \makecellslrableft{Subsequent A/B test\\ execution} & The results of the A/B test are used to execute subsequent A/B tests, e.g. using the results of A/B tests to automatically determine which subsequent A/B tests to execute~\cite{primary-study-2145}. & 4 \\
    
    Unspecified & The use of the test results was not specified in the study. & 24 \\
    \bottomrule
\end{tabular}
\label{tab:use_of_test_results}
\end{altrowcolortbl}

\subsubsection{Role of stakeholders}

To address the second part of RQ3 (What is the role of stakeholders in this process?), we analyze the role of stakeholders in A/B test execution. 

\paragraph{Roles of stakeholders} %

Table~\ref{tab:roh_execution} lists the different role of stakeholders in the A/B test execution we have extracted from the primary studies with associated tasks, a description and examples. 
We identified two main roles: experiment contributor (40 occurrences) and experiment assessor (37 occurrences). The role \emph{Experiment contributor} consists of managing the A/B test execution. The role \emph{Experiment assessor} consists of evaluating the A/B test results and potentially undertaking additional actions. The top task of the experiment contributor is experiment supervision (19 occurrences). The top task of the experiment assessor is experiment post-analysis (17 occurrences).

\begin{altrowcolortbl}
\centering
\caption{Identified roles and concrete tasks of stakeholders during in the execution of A/B tests.}
\renewcommand{\arraystretch}{1.5}
\begin{tabular}{M{.8cm}m{0.1cm}@{}m{4.0cm}m{7.3cm}c}
    \toprule
    \textbf{Role} & & \textbf{Task} & \textbf{Task description} & \textbf{Occ.} \\\midrule
    \cellcolor{white} & \cellcolor{gray!80} & Experiment supervision & Monitoring and closely following up on the execution of A/B tests~\cite{primary-study-2145, primary-study-143}. & 19 \\
    
    \cellcolor{white} & \cellcolor{gray!80} & Experiment alteration & Altering aspects of the A/B test during its execution. Examples include adjusting the population assignment of the experiment~\cite{primary-study-278}, or adjusting the A/B variants themselves~\cite{primary-study-2275}. & 12 \\
    
    \cellcolor{white} \multirow{-3}{*}[68pt]{\rotatebox[origin=c]{90}{\textbf{Experiment contributor (40)}}} & \cellcolor{gray!80} & Experiment termination &Stopping A/B tests when deemed necessary. Examples include manually stopping A/B tests when sufficient data is collected~\cite{primary-study-2055}, or stopping the experiment early when harm is observed~\cite{primary-study-16}. & 9 \\

    \cellcolor{white} & \cellcolor{gray!30} & Experiment post-analysis & Various steps that are taken after analyzing the results of the A/B test. Examples include double checking results from executed A/B tests~\cite{primary-study-699}, performing a deeper analysis of suspicious results~\cite{primary-study-31}, or performing bias reduction techniques on the retrieved data from the A/B tests~\cite{primary-study-770}. & 17 \\
    
    \cellcolor{white} & \cellcolor{gray!30} & Experiment triggering & Starting the execution of (subsequent) A/B tests~\cite{primary-study-615, primary-study-2042}. & 13 \\
    
    \cellcolor{white} \multirow{-3}{*}[63pt]{\rotatebox[origin=c]{90}{\textbf{Experiment assessor (37)}}} & \cellcolor{gray!30} & Other & This category encompasses a few niche tasks, such as documenting the findings and learning from conducting the A/B test~\cite{primary-study-34}, rerunning A/B tests~\cite{primary-study-100}, or incorporating user feedback in the analysis of the A/B tests~\cite{primary-study-115}. & 7 \\ 
    \bottomrule
\end{tabular}
\label{tab:roh_execution}
\end{altrowcolortbl}

\subsubsection{Cross analysis A/B test execution}

We take a deeper look at two mappings of data items related to the execution of A/B tests: Use of test results with the tasks of stakeholders in the execution of A/B tests; and the evaluation method with the tasks of stakeholders in the execution of A/B tests.

\paragraph{Use of test results vs Tasks of stakeholders in the execution of A/B tests}

The first mapping we analyze relates to the use of test results and the tasks stakeholders undertake in the execution of A/B tests. The results are shown in Table~\ref{tab:use_of_test_results_role_of_humans_in_the_experiment_execution}. We highlight some key observations:

\begin{itemize}
    \item Experiment supervision is applied regardless of the usages of test results. For feature rollout as a use of A/B test results, the task of experiment supervision is often mentioned. Supervision takes on a key task in this context to ensure that the rollout happens in a hazard-free manner (i.e., no harm is caused to users)~\cite{primary-study-615, primary-study-12}.
    
    \item The task of experiment post-analysis is typically only reported for experiments that are fully complete (i.e., do not go through additional rounds of iteration). In the primary studies where the results of the A/B tests are used for subsequent A/B test design, no instances were identified where stakeholders take the task of performing post-analysis on the results of the experiments.   

    \item For subsequent A/B test design, the task of experiment triggering is often mentioned. This is to be expected since the newly designed A/B tests also need to be executed. Additionally, A/B test termination is also mentioned often (e.g., terminating an experiment due to bad results~\cite{primary-study-695}).
    
    \item In the case of bug fixing and detection, stakeholders typically supervise experiments (either to detect possible bugs in the code or ensure the bugfix is effective)~\cite{primary-study-747}, and trigger the experiments (i.e. launch an experiment explicitly to fix a known bug in the application)~\cite{primary-study-60}. 
    
\end{itemize}

\begin{table}
\centering
\caption{Use of test results × Tasks of stakeholders in the experiment execution ("cont. feature dev." is short for "continue feature development", "val. eff." is short for "validation of effectiveness", and "val. of a RQ" is short for "validation of a research question"). }
\begin{tabular}{lcccccc}
    \toprule
    \customDiagbox[width=\widthof{\textbf{Val. eff. A/B testing}}]{Use}{Task} & \makecellslrabbold{Supervision} & \makecellslrabbold{Post-\\analysis} & \makecellslrabbold{Triggering} & \makecellslrabbold{Alteration} & \makecellslrabbold{Termination}
    \\\midrule
    
    \textbf{Feature selection}     & 8  & 11 & 6 & 8 & 4 
    \\
    \textbf{Feature rollout}       & 10 & 4  & 6 & 6 & 4 
    \\
    \textbf{Cont. feature dev.}    & 7  & 3  & 5 & 2 & 3 
    \\
    \textbf{A/B test design}       & 6  & 0  & 5 & 2 & 3  
    \\
    \textbf{Val. eff. A/B testing} & 1  & 2  & 1 & 1 & 1  
    \\
    \textbf{Val. of a RQ}          & 1  & 1  & 0 & 1 & 1  
    \\
    \textbf{Bug detection/fixing}  & 4  & 0  & 3 & 2 & 2 
    \\
    \textbf{A/B test execution}    & 1  & 0  & 1 & 0 & 0 
    \\

    \bottomrule
\end{tabular}
\label{tab:use_of_test_results_role_of_humans_in_the_experiment_execution}
\end{table}

\paragraph{Evaluation method vs Tasks of stakeholders in the execution of A/B tests}

In addition, we analyze the tasks stakeholders undertake during the execution of A/B tests across the evaluation methods. This mapping is shown in Table~\ref{tab:evaluation_method_role_of_humans_in_the_experiment_execution}. We highlight a number of key takeaways:

\begin{itemize}
    
    \item All tasks that stakeholders undertake in the execution of A/B tests are widely encountered in the case of empirical evaluation.
    
    \item For the method of simulation based on real empirical data, the task of post-analysis is reported more often than any other task. An example is looking for outliers in the analysis of the results of A/B tests, and using historical experiments to confirm its effectiveness~\cite{primary-study-804}.
    
    \item Primary studies that use simulation as an evaluation method rarely specify the tasks stakeholders undertake in the execution of A/B tests. We hypothesize that, since simulations allow for a more controlled way of conducting A/B tests, the tasks stakeholders undertake after the design of A/B tests are not pertinent. 

    \item The only stakeholder task reported for theoretical evaluation is experiment alteration (primary study~\cite{primary-study-209}). 
\end{itemize}

\begin{table}
\centering
\caption{Evaluation method × Tasks of stakeholders in the test execution ("emp. sim." short for "simulation based on real empirical data", "ill." short for "illustrative").}
\begin{tabular}{lcccccc}
    \toprule
    \customDiagbox[width=\widthof{\textbf{Ill. example}}]{Method}{Task} & \makecellslrabbold{Supervision} & \makecellslrabbold{Post-\\analysis} & \makecellslrabbold{Triggering} & \makecellslrabbold{Alteration} & \makecellslrabbold{Termination} & \makecellslrabbold{Other} \\\midrule
    
    \textbf{Empirical}      & 14 & 13 & 10 & 10 & 6 & 6 \\
    \textbf{Emp. sim.}      & 2  & 4  & 1  & 0  & 1 & 0 \\
    \textbf{Simulation}     & 1  & 1  & 1  & 0  & 0 & 0 \\
    \textbf{Ill. example}   & 2  & 0  & 1  & 2  & 2 & 1 \\
    \textbf{Case study}     & 0  & 0  & 0  & 0  & 0 & 0 \\
    \textbf{Theoretical}    & 0  & 0  & 0  & 1  & 0 & 0 \\

    \bottomrule
\end{tabular}
\label{tab:evaluation_method_role_of_humans_in_the_experiment_execution}
\end{table}

\begin{tcolorbox}[colback=white]
\textbf{Research Question 3: How are A/B tests executed in the system? What is the role of stakeholders in this process?} %
The main types of data collected during the A/B test execution relate to the product/system, users, and spatial-temporal aspects. The dominating evaluation method used in A/B testing is empirical evaluation, but a relevant number of studies also use simulation. A/B test results are primarily used for feature selection, followed by feature rollout, and continue feature development. (Automatic) subsequent A/B test execution is only used marginally. The main reported roles of stakeholders in A/B test execution is experiment contributor (with experiment supervisor as main task) and experiment assessor (with experiment post-analysis as main task).
\end{tcolorbox}

\subsection{RQ4: What are the reported open research problems in the field of A/B testing?}

To answer research question 4, we analyze the results of data item Open problems (\textbf{D18}).

Table~\ref{tab:open_problems} present a categorization of open problems we have identified in the primary studies. For each category we devised concrete sub-categories of open problems. We elaborate on each type of open problem with illustrative examples.

\begin{table}
\centering
\caption{List of identified open problems.}
\begin{tabular}{llc}
    \toprule
    \makecellslrableft{\textbf{Open problem}\\\textbf{category}} & \textbf{Open problem sub-category} & \occurrences{} \\\midrule 
    & Extend the evaluation & 21 \\
    & Provide thorough analysis of approach & 16 \\
    \multirow{-3}{*}{Evaluation-related} & Other evaluation-related & 34 \\[8pt]
    & Add process guidelines & 9 \\
    \multirow{-2}{*}{Process-related} & Automate process & 7 \\[8pt]
    & Enhance scalability & 7 \\
    \multirow{-2}{*}{Quality-related} & Enhance applicability & 6 \\
    \bottomrule
\end{tabular}
\label{tab:open_problems}
\end{table}

\subsubsection{Evaluation-related open problems}

First, we established three sub-categories of open problems that are related to the evaluation of the proposed approach: (1) extensions to the evaluation of the approach presented in the primary study, (2) a more thorough analysis of the approach presented in the primary study, and (3) Other evaluation-related open problems in the primary study.

\paragraph{Extend the evaluation} %
Drutsa et al.~\cite{primary-study-56} explore periodicity patterns in user engagement metrics, and its influence on engagement metrics in A/B tests. Moreover, the authors put forward new A/B metrics that take such periodicity patterns into account, resulting in more sensitive A/B test analysis. The authors evaluated the proposed metrics on historical A/B test data from Yandex, though they state that further evaluation of the approach could be carried out in different domains such as social networks, email services, and video/image hosting services.
From a slightly different point of view, Barajas et al.~\cite{primary-study-198} developed a technique to determine the causal effects of marketing campaigns on users, putting the focus on the campaign itself rather than only focusing on the design of advertisement media. The authors put forward specific guidelines on randomizing and assigning users to advertising campaigns, and provide a technique to estimate the causal effect the campaigns have on the users under test. As a point of future work, the authors posit a different evaluation question concerning what would have happened if the technique would have been applied to the whole population.

\paragraph{Provide thorough analysis of approach} %
An example of this category is mentioned by Peska and Vojtas~\cite{primary-study-561}. The authors put forward a way of evaluating recommendation algorithms in small e-commerce applications both offline and online via A/B testing. The approach compares results of offline evaluation of recommendation algorithms with the results of online A/B testing of the algorithms. Moreover, the authors then used these data to build a prediction model to determine the promising recommendation algorithms more effectively due to the knowledge obtained from online A/B testing. As future work, the authors list that further work is necessary to verify the causality of an effect observed in the analysis of offline and online A/B testing data.
In another primary study written by Madlberger and Jizdny~\cite{primary-study-1453}, the authors perform an analysis on the impact of social media marketing on click-through rates and customer engagement. To accomplish this, they run multiple social media marketing campaigns using A/B testing, evaluating hypotheses related to the impact of visual and content aspects of advertisements on the click rates of end-users. As future research, the authors report that a more comprehensive investigation is necessary to ascertain why some hypotheses in the study have been rejected.

\paragraph{Other evaluation-related} %
An example of other evaluation-related open problems is laid out by Gruson et al.~\cite{primary-study-292}. The authors propose a methodology based on counterfactual analysis to evaluate recommendation algorithms, leveraging both offline evaluation and online evaluation via A/B testing. The approach comprises A/B testing recommendations to a subset of the population, and using the results of these tests to de-bias offline evaluations of the recommendation algorithm based on historical data. In regards to open problems, the authors mention exploring additional metrics for the approach, as well as potential improvements that can be made to the estimators they use in the approach.
Another example is specified by Ju et al.~\cite{primary-study-89}, who present an alternative to standard A/B testing with a static hypothesis test by putting forward a sequential test. Classically in A/B testing, the hypothesis of the test is tested after a fixed time and conclusions are made based on the final result. The sequential test put forward by the authors does not have a predetermined number of observations, rather at multiple points during the experiment the test determines whether the hypothesis can be accepted, rejected, or if more observations are required. For future work, the authors wish to support A/B/n experiments in their approach, as well as extending the procedure for data that follows a non-binomial distribution.
In a final example, Gui et al.~\cite{primary-study-21} study ways of dealing with interference of network effects in the results of A/B tests. One of the fundamental assumptions of A/B testing is that users are only affected by the A/B variant they are assigned to. However, network effects can undermine this assumption do to interaction between users in the population. The authors demonstrate the presence of network effects at LinkedIn, and propose an estimator for the average treatment effect that also takes potential network effects into account. As a line of future research, the authors want to investigate ways of enhancing the approach such that it can deal with more real life phenomena.

\subsubsection{Process-related open problems}

Second, we established two sub-categories of open problems that are process-related: (1) guidelines to the A/B testing process, and (2) automation of aspects of the A/B testing process.

\paragraph{Add process guidelines} %
In an effort to provide more nuanced A/B testing guidelines in the e-commerce domain, Goswami et al.~\cite{primary-study-699} discuss controlled experiments to make decisions in the context of e-commerce search. %
Considerations such as how to prioritize projects for A/B testing for smaller retailers and how to conduct A/B tests during holiday time are left as open questions.
A different primary study covering the benefits of controlled experimentation at scale is presented by Fabijan et al.~\cite{primary-study-747}. In this study, the authors present multiple examples of conducted A/B tests, and the corresponding lessons learned from these experiments. %
One of the listed open problems in the study relates to providing "guidance on detection of patterns between leading and lagging metrics".

\paragraph{Automate process} \label{sec:openproblems:automation}%
Mattos et al.~\cite{primary-study-1159} present a step towards automated continuous experimentation.
The authors put forward an architectural framework that accommodates the automated execution of A/B tests and automated generation of A/B variants. 
To validate the framework, an A/B test was conducted with a robot. One of the open challenges laid out in the study comprises the ability to automatically generate hypotheses for A/B tests based on the collected data.
Duivesteijn et al.~\cite{primary-study-1965} present A\&B testing, an approach that leverages exceptional model mining techniques to target A/B variants to subgroups in the population under test. As opposed to deploying the best-performing variant of the A/B test, the authors put forward running both variants (if ample resources are available) and targeting specific variants to individual users based on their inferred subgroups. One of the potential avenues for future research consists of the development of a framework that would enable automated personalization of websites supported by A/B testing.

\subsubsection{Quality-related open problems}

Lastly, we established two sub-categories of open problems that are quality-related: (1) enhancing scalability of the proposed approach, and (2) enhancing the applicability of the approach.

\paragraph{Enhance scalability} %
One example of this is presented by Zhao et al.~\cite{primary-study-835}. In order to obtain a causal explanation behind the results of A/B tests, the authors propose segmenting the population, and consequently analyzing the results of the A/B test in individual segments. For future work, the authors mention developing a more scalable solution that integrates the approach into their existing experimentation platform.
To address online experimentation specifically for cloud applications, Toslali et al.~\cite{primary-study-1623} introduce Jackpot, a system for online experimentation in the cloud. Jackpot supports multivariate A/B testing and ensures proper management of interactions in the cloud application during the execution of A/B tests. As a venue for future work, the authors mention ways of dealing with the limited scalability of multivariate experimentation due to the number of potential experiments increasing exponentially with the number of elements to be tested.

\paragraph{Enhance applicability} %
One such study explores A/B testing in the automotive industry~\cite{primary-study-682}. The study addresses concerns relating to the limited sample sizes A/B tests obtain due to the limited nature of participants that can take part in A/B tests in the industry. To overcome this hurdle, the authors provide specific guidelines for performing A/B testing and determining the assignment of users to either the control or treatment variant in the test. However, one limitation pertains to requiring pre-experimental data to ensure a balanced population assignment between both A/B variants.
In an effort to increase sensitivity in A/B testing, Liou and Taylor~\cite{primary-study-77} propose a new estimator for A/B testing that takes variance of individual users into account. To realize this, pre-experiment data of individual users is analyzed and variances are computed. In order to validate the approach a sample of 100 previously conducted A/B tests were collected and analyzed using the new approach. A big limitation noted by the authors is that "a stronger assumption about the homogeneity of the treatment effect" is required in order for the approach to remain unbiased.

\begin{tcolorbox}[colback=white]
\textbf{Research Question 4: What are the reported open research problems in the field of A/B testing?} %
The most commonly reported open problems directly related to the proposed approach, in particular improving the approach, extending the approach and providing a thorough analysis. Other less frequently reported open problems relate to the A/B testing process, in particular adding guidelines for the A/B testing process, and automating the process. Finally, a number of studies report open problems regarding quality properties, specifically enhancing scalability and applicability of the proposed approach.
\end{tcolorbox}

\section{Discussion}\label{sec:discussion} 

In this section, we discuss a number of additional insights we obtained. We start with the research topics studied by the primary studies. Next, we look at environments and tools used for A/B testing. Then we report a number of opportunities for future research. We conclude with a discussion of threats to validity of the study. 

\subsection{Research topics}

During data extraction of the \numberofpapers{} primary studies, we noted the general subject matters of the primary studies and categorized the primary studies along 7 research topics. Table~\ref{tab:categorizations} summarizes these 7 topics. Note that studies share overlapping topics. We briefly explain now each category and provide a few examples from the primary studies.

\begin{altrowcolortbl}
\centering
\caption{Research topics of primary studies.}
\renewcommand{\arraystretch}{1.3}
\begin{tabular}{lcm{6.4cm}}
    \toprule
    \textbf{Topic} & \occurrences{} & \textbf{Primary studies} \\\midrule
    Application of A/B testing & 51 & \cite{primary-study-105, primary-study-125, primary-study-139, primary-study-189, primary-study-196, primary-study-198, primary-study-209, primary-study-231, primary-study-244, primary-study-277, primary-study-278, primary-study-294, primary-study-321, primary-study-350, primary-study-351, primary-study-355, primary-study-360, primary-study-362, primary-study-436, primary-study-459, primary-study-473, primary-study-476, primary-study-482, primary-study-506, primary-study-507, primary-study-569, primary-study-601, primary-study-614, primary-study-636, primary-study-694, primary-study-707, primary-study-727, primary-study-760, primary-study-763, primary-study-785, primary-study-1312, primary-study-1337, primary-study-1453, primary-study-1476, primary-study-1572, primary-study-1718, primary-study-1733, primary-study-1956, primary-study-2026, primary-study-2042, primary-study-2060, primary-study-2066, primary-study-2088, primary-study-2287, snowballed-study-1, snowballed-study-4} \\
    
    Improving efficiency of A/B testing & 20 & \cite{primary-study-1, primary-study-12, primary-study-14, primary-study-17, primary-study-27, primary-study-38, primary-study-42, primary-study-45, primary-study-48, primary-study-52, primary-study-54, primary-study-56, primary-study-67, primary-study-77, primary-study-82, primary-study-89, primary-study-693, primary-study-804, snowballed-study-2, snowballed-study-3} \\
    
    Beyond standard A/B testing & 18 & \cite{primary-study-23, primary-study-26, primary-study-36, primary-study-47, primary-study-55, primary-study-68, primary-study-100, primary-study-245, primary-study-292, primary-study-306, primary-study-561, primary-study-709, primary-study-1159, primary-study-1636, primary-study-1965, primary-study-1966, primary-study-1970, primary-study-2145} \\
    
    Concrete A/B testing problems & 17 & \cite{primary-study-7, primary-study-21, primary-study-44, primary-study-60, primary-study-91, primary-study-143, primary-study-145, primary-study-327, primary-study-668, primary-study-682, primary-study-699, primary-study-1495, primary-study-1623, primary-study-1679, primary-study-2055, primary-study-2215, primary-study-2293} \\
    
    Pitfalls and challenges of A/B testing & 13 & \cite{primary-study-10, primary-study-15, primary-study-25, primary-study-31, primary-study-51, primary-study-80, primary-study-84, primary-study-676, primary-study-697, primary-study-720, primary-study-770, primary-study-1937, primary-study-2161} \\
    
    Experimentation frameworks and platforms & 13 & \cite{primary-study-34, primary-study-94, primary-study-115, primary-study-119, primary-study-169, primary-study-174, primary-study-227, primary-study-293, primary-study-679, primary-study-680, primary-study-835, primary-study-1587, primary-study-2275} \\
    
    A/B testing at scale & 9 & \cite{primary-study-16, primary-study-28, primary-study-50, primary-study-615, primary-study-681, primary-study-683, primary-study-695, primary-study-747, primary-study-1796} \\
     \bottomrule
\end{tabular}
\label{tab:categorizations}
\end{altrowcolortbl}

\subsubsection{Application of A/B testing}

The main focus of the primary study is the use and application of A/B testing as evaluation tool for the main subject matter of the study (e.g. evaluation new recommendation algorithm, interface redesigns, etc\footnote{See data item \textit{A/B target} in Section~\ref{sec:ab-target} for specific references.}).

\subsubsection{Improving the efficiency of A/B testing}

This topic is about improving the process of A/B testing by exploring ways of improving sensitivity in A/B testing data~\cite{primary-study-48, primary-study-14, primary-study-17, primary-study-38}, investigating sequential testing techniques to stop A/B tests as soon as reasonable~\cite{primary-study-89, primary-study-52, primary-study-1}, proposing techniques to detect invalid A/B tests\footnote{Invalid refers to badly designed experiments or misinterpretation of the results retrieved from the experiment.}~\cite{primary-study-12}, and using extra data such as periodicity patterns in user behavior to improve A/B testing~\cite{primary-study-56}.

\subsubsection{Beyond standard A/B testing} \label{subsubsec:beyondstandard}

This topic is about techniques that go beyond standard A/B testing, such as the use of new types of A/B metrics~\cite{primary-study-23, primary-study-47, primary-study-100}, use of counterfactuals in the evaluation of A/B tests\footnote{Counterfactual analysis provides answers to the cause and effect of the treatment group and their corresponding outcomes, compared to what would have happened if the treatment would not have been applied.}~\cite{primary-study-55, primary-study-245}, investigating ways of automating parts of the A/B testing process~\cite{primary-study-306, primary-study-1159, primary-study-1966, primary-study-2145}, improving or altering the A/A testing process~\cite{primary-study-68, primary-study-709},
and investigating ways of combining offline- and online A/B testing~\cite{primary-study-292, primary-study-561}.

\subsubsection{Concrete A/B testing problems}

This topic includes studies that A/B testing in specific  domains and specific types of A/B testing. Examples include A/B testing specifically in the e-commerce domain~\cite{primary-study-2055, primary-study-699}, network A/B testing or A/B testing in marketplaces~\cite{primary-study-668, primary-study-21, primary-study-1495},  A/B testing in the CPS domain with digital twins~\cite{primary-study-143}, or A/B testing for mobile applications~\cite{primary-study-2215, primary-study-44}.

\subsubsection{Pitfalls and challenges of A/B testing}

This topic is about pitfalls related to conducting A/B testing~\cite{primary-study-25, primary-study-15, primary-study-80, primary-study-720}, or (particular domain-related) challenges related to A/B testing~\cite{primary-study-84, primary-study-676, primary-study-770}.

\subsubsection{Experimentation frameworks and platforms}

This topic covers papers that present an A/B testing platform~\cite{primary-study-115, primary-study-2275, primary-study-34}, or a framework concerning aspects related to the A/B testing process such as a framework for detecting data loss in A/B tests~\cite{primary-study-293}, a framework for the design of A/B tests~\cite{primary-study-680}, or a framework for personalization of A/B testing~\cite{primary-study-94}.

\subsubsection{A/B testing at scale}

Primary studies under this topic focus on conducting A/B testing at a large scale, e.g., considerations for conducting A/B testing at scale~\cite{primary-study-681, primary-study-683, primary-study-695}, process models or guidelines for A/B testing at scale~\cite{primary-study-50, primary-study-615}, or concrete scalable solutions such as a scalable statistical method for measuring quantile treatment effects for performance metrics in A/B tests~\cite{primary-study-28}.

\subsection{Environments and tools used for A/B testing}

In addition to the research topics covered in the primary studies, we also analyze the environments and tools that were used to realize A/B testing, see Table~\ref{tab:environments_ab_testing}.

\begin{table}
\caption{Environments and tools used for A/B testing.}
\centering
\begin{tabular}{lc}
    \toprule
    \textbf{Environment} & \occurrences{} \\\midrule
    In-house experimentation system & 20 \\
    Research tool or prototype & 13 \\
    Commercial A/B testing tool & 10 \\
    Commercial non A/B testing tool & 7 \\
    User survey & 1 \\
    \bottomrule
\end{tabular}
\label{tab:environments_ab_testing}
\end{table}

The most commonly mentioned type of environment is in-house experimentation system for A/B testing (20 occurrences), for instance dedicated environments developed by companies such as Microsoft~\cite{primary-study-60}, Google~\cite{primary-study-2275}, eBay~\cite{primary-study-683}, and Etsy~\cite{primary-study-89}. These environments broadly support executing A/B tests. Furthermore, some primary studies describe concrete features of the experimentation system to help design A/B tests, e.g. controlling for bias during the specification of A/B tests in Airbnb's Experimentation Reporting Framework~\cite{primary-study-82}.
Next, we observe research tools and prototypes (13 occurrences). Examples include a tool to perform online cloud experimentation~\cite{primary-study-1623}, a research prototype for A/B testing implemented in NodeJS~\cite{primary-study-306}, a tool for A/B testing with decision assistants~\cite{primary-study-2055}, and a tool that enables automatic execution of multiple A/B tests~\cite{primary-study-2145}.
The remaining environments we identified were commercial A/B testing tools (10 occurrences), e.g., Optimizely~\cite{primary-study-1572}, and Google Analytics~\cite{primary-study-2042}; commercial tools not related to A/B testing (7 occurrences), e.g., Crazy egg~\cite{primary-study-2042}, a heatmapping tool used to design A/B variants, and using Yahoo Gemini (advertisement platform) to test different advertising strategies~\cite{primary-study-1453}; and a user survey (1 occurrence) to determine which A/B variants to test by conducting a preliminary survey.

\subsection{Research opportunities and future research directions}

From our study, we propose a number of potential future research directions in the field of A/B testing. Concretely, we provide three lines of research: research on further improving the general process of A/B testing, research on automating aspects of A/B testing, and research on the adoption of proposed statistical methods in A/B testing.

\subsubsection{Improving the A/B testing process}

One future direction relates to taking considerations when running many A/B tests at once~\cite{primary-study-2275}. Plenty of studies cover this topic by e.g., discussing lessons learned in unexpected A/B test results that were caused by other A/B tests that were running in parallel~\cite{primary-study-25}, or manually checking for possible effects of running A/B tests by analyzing the deployed A/B tests in the system~\cite{primary-study-683}. Yet, we did not encounter a study that puts forward a systematic approach to tackle this problem.

Another avenue for future research is about improving the sensitivity in A/B tests by, e.g., combining different sensitivity improvement techniques as pointed out by Drutsa et al.~\cite{primary-study-48}, enabling proactive prediction of user behavior in A/B tests based on historical data~\cite{primary-study-56}, and a deeper study of A/B test estimators to achieve better sensitivity as mentioned by Poyarkov et al.~\cite{primary-study-14}.

The last avenue for future research in improving the A/B testing process relates to providing further guidelines and designing principles for choosing and engineering A/B metrics. We highlight two primary studies that mention open problems related to this opportunity: Kharitonov et al.~\cite{primary-study-38} put forward learning sensitive combinations of A/B metrics as a general open problem, and Duan et al.~\cite{primary-study-47} discuss investigating dynamics between surrogate metrics and the actual underlying metric.

\subsubsection{Automation}

In an effort to establish continuous experimentation, multiple studies put forward steps companies can take to develop an experimentation culture, e.g.~\cite{primary-study-50, primary-study-84, Fabijan2021}. 
In light of expanding this experimentation culture, (partial) automation of the A/B testing process is essential to enable and empower continuous experimentation~\cite{primary-study-12, primary-study-699}. Initial research on automation of steps in the A/B testing has been conducted, as for example presented by Tamburrelli et al~\cite{primary-study-2145} and Mattos et al.~\cite{primary-study-1159}, see Sections~\ref{subsubsec:beyondstandard} and~\ref{sec:openproblems:automation}. Yet the present state of research in this topic suggests that further investigation and more in-depth solutions are necessary to fully exploit automated design and execution of A/B tests.
Additionally, a number of open problems still remain that could facilitate and enable automated experimentation, e.g., determining which A/B tests to prioritize at execution~\cite{primary-study-699}, and automatically generating insights related to the rationale and cause of experiment results to experiment developers to guide product development~\cite{primary-study-84}.

\subsubsection{Adoption and tailoring statistical methods}

Even though a number of primary studies discuss bootstrapping as a technique to evaluate the results of A/B tests~\cite{primary-study-94, primary-study-569, primary-study-699}, bootstrapping remains largely unexplored in A/B testing, despite the fact that this statistical method has the potential to improve the analysis of A/B test results~\cite{primary-study-681, primary-study-327}. Moreover, bootstrapping can present an invaluable tool to provide statistical insights into the results of the tests which could e.g. not be obtained by a standard equality testing method~\cite{Efron1994}. However, one big downside of bootstrapping is that it is computationally expensive~\cite{primary-study-770}.
Alongside adoption of known statistical methods, designing and tailoring new statistical methods to accommodate for particular experimentation scenarios presents an interesting research direction. One example is mentioned by Kharitonov~\cite{primary-study-52}, who put forward designing a custom statistical test for non-binomial A/B metrics. Another example concerns taking into account "the effects from multiple treatments with various metrics of interest" to tailor the approach presented by Tu et al.~\cite{primary-study-94} for optimal treatment assignments in A/B testing by leveraging causal effect estimations.

Besides a limited number of primary studies employing bootstrapping in the analysis of A/B tests, a significant number of studies mention statistically significant results or p-values in the analysis of conducted A/B tests without specifying the concrete statistical test used (37 occurrences). Moreover, a considerable number of studies do not report anything related to statistical analysis (47 occurrences). We argue that this information is important to report in research publications, and urge authors to specify the concrete statistical methods used\footnote{Or alternatively an explicit mention of lack of statistical methods used.} to obtain the results in the studies.

\subsection{Threats to validity}
In this section we list potential threats to the validity of the systematic literature review~\cite{Ampatzoglou2019}.

\subsubsection{Internal validity}

Internal validity refers to the extent to which a causal conclusion based on a study is warranted.
One threat to the internal validity is a potential bias of researchers that perform the SLR, which may have an effect on the data collection and the insights derived in the study. In order to mitigate this threat, we involved multiple researchers in the study. Multiple researchers were responsible for selecting papers, extracting data and analyzing results. In each step, cross-checking was applied to minimize bias. Extra researchers were involved if no consensus could be found. Additionally, we defined a rigid protocol for the systematic literature review.

\subsubsection{External validity}

External validity refers to the extent to which the findings of the study can be generalized to the general field of A/B testing.
A threat to the external validity of this systematic literature review is that not all relevant works are covered. To mitigate this threat, we searched all main digital library sources that publish work in computer science. Secondly, we defined the search string by including all commonly used terms for A/B testing to ensure proper retrieval of relevant works. Lastly, we also applied snowballing on the selected papers from the automatic search query to uncover additional works that might have been missed.

\subsubsection{Conclusion validity}

Conclusion validity refers to the extent to which we obtained the right measure and whether we defined the right scope in relation to what is considered research in the field of A/B testing.
One threat to the conclusion validity is the quality of the selected studies; studies of lower quality might produce insights that are not justified or applicable to the general field of A/B testing. In order to mitigate this threat, we excluded short papers, demo papers, and roadmap papers from the study. Furthermore, we evaluated a quality score for each selected paper. Papers with a quality score $\leq 4$ were excluded from the study.

\subsubsection{Reliability}

Reliability refers to the extent to which this work is reproducible if the study would be conducted again.
To mitigate this threat, we make all the collected and processed data available online. We also defined a specific search string, a list of online sources, and other specific details in the research protocol to ensure reproducibility. Bias of researchers also poses a threat here, influencing that similar results would be retrieved if the systematic literature review would be conducted again with a different set of reviewers.

\section{Conclusion}\label{sec:conclusions}

A/B testing supports data-driven decisions about the adoption of features. It is widely used across different industries and key technology companies such as Google, Meta, and Microsoft. In this systematic literature review, we identified the subjects of A/B tests, how A/B tests are designed and executed, and the reported open research problems in the literature. 
We observed that algorithms, visual elements, and changes to a workflow or process are most commonly tested, with web, search engine, and e-commerce being the most popular application domains for A/B testing. %
Concerning the design of A/B tests, classic A/B tests with two variants are most commonly used, alongside engagement metrics such as conversion rate or number of impressions as metric to gauge the potential of the A/B variants. Hypothesis tests for equality testing are broadly utilized to analyze A/B test results, and bootstrapping also garners interest in a few primary studies. We devised three roles stakeholders take on in the design of A/B tests: Concept designer, Experiment architect, and Setup technician. %
Regarding the execution of A/B tests, empirical evaluation is the leading evaluation method. Besides the main A/B metrics, data concerning the product or system, and user-centric data are collected the most to conduct deeper analysis of the results of the A/B tests. A/B testing is most commonly used to determine and deploy the better performing A/B variant, or to gradually roll out a feature. Lastly, we devised two roles stakeholders take on in the execution of A/B tests: Experiment contributor, and Experiment assessor.

We identified seven categories of open problems: improving proposed approaches, extending the evaluation of the proposed approach, providing thorough analysis of the proposed approach, adding A/B testing process guidelines, automating the A/B testing process, enhancing scalability, and enhancing applicability. Leveraging these categories and observations made during the analysis, we provide three main lines of interesting research opportunities: developing more in-depth solutions to automate stages of the A/B testing process; presenting improvements to the A/B testing process by examining promising avenues for sensitivity improvement, systematic solutions to deal with interference of many A/B tests running at once, and providing guidelines and designing principles to choose and engineer A/B metrics; and lastly the adoption and tailoring of more sophisticated statistical methods such as bootstrapping to strengthen the analysis of A/B testing further.

\section*{Acknowledgement}
We thank Michiel Provoost for his support to this study.

\bibliographystyle{ACM-Reference-Format}
\bibliography{ref/main, ref/primary-studies, ref/snowball}

\newpage
\appendix

\section{List of primary studies}

\begingroup
\rowcolors{2}{white}{gray!25}
\begin{longtable}{M{1.3cm}M{1.75cm}m{12cm}}
\caption{List of primary studies.} 
\\
    \rowcolor{gray!50}
    \textbf{Paper ID} & \textbf{Reference} & \textbf{Title} \\
    1 & \cite{primary-study-1} & A Nonparametric Sequential Test for Online Randomized Experiments \\
    2 & \cite{primary-study-7} & Detecting Network Effects: Randomizing Over Randomized Experiments \\
    3 & \cite{primary-study-10} & Unexpected Results in Online Controlled Experiments \\
    4 & \cite{primary-study-12} & How A/B Tests Could Go Wrong: Automatic Diagnosis of Invalid Online Experiments \\
    5 & \cite{primary-study-14} & Boosted Decision Tree Regression Adjustment for Variance Reduction in Online Controlled Experiments \\
    6 & \cite{primary-study-15} & Trustworthy Online Controlled Experiments: Five Puzzling Outcomes Explained \\
    7 & \cite{primary-study-16} & Online Controlled Experiments at Large Scale \\
    8 & \cite{primary-study-17} & Non-Stationary A/B Tests \\
    9 & \cite{primary-study-21} & Network A/B Testing: From Sampling to Estimation \\
    10 & \cite{primary-study-23} & False Discovery Rate Controlled Heterogeneous Treatment Effect Detection for Online Controlled Experiments \\
    11 & \cite{primary-study-25} & Experimentation Pitfalls to Avoid in A/B Testing for Online Personalization \\
    12 & \cite{primary-study-26} & Statistical Inference in Two-Stage Online Controlled Experiments with Treatment Selection and Validation \\
    13 & \cite{primary-study-27} & Consistent Transformation of Ratio Metrics for Efficient Online Controlled Experiments \\
    14 & \cite{primary-study-28} & CONQ: CONtinuous Quantile Treatment Effects for Large-Scale Online Controlled Experiments \\
    15 & \cite{primary-study-31} & Diagnosing Sample Ratio Mismatch in Online Controlled Experiments: A Taxonomy and Rules of Thumb for Practitioners \\
    16 & \cite{primary-study-34} & IPEAD A/B Test Execution Framework \\
    17 & \cite{primary-study-36} & Peeking at A/B Tests: Why It Matters, and What to Do about It \\
    18 & \cite{primary-study-38} & Learning Sensitive Combinations of A/B Test Metrics \\
    19 & \cite{primary-study-42} & Practical Aspects of Sensitivity in Online Experimentation with User Engagement Metrics \\
    20 & \cite{primary-study-44} & Evaluating Mobile Apps with A/B and Quasi A/B Tests \\
    21 & \cite{primary-study-45} & On Post-Selection Inference in A/B Testing \\
    22 & \cite{primary-study-47} & Online Experimentation with Surrogate Metrics: Guidelines and a Case Study \\
    23 & \cite{primary-study-48} & Future User Engagement Prediction and Its Application to Improve the Sensitivity of Online Experiments \\
    24 & \cite{primary-study-50} & The Evolution of Continuous Experimentation in Software Product Development: From Data to a Data-Driven Organization at Scale \\
    25 & \cite{primary-study-51} & How to Measure Your App: A Couple of Pitfalls and Remedies in Measuring App Performance in Online Controlled Experiments \\
    26 & \cite{primary-study-52} & Sequential Testing for Early Stopping of Online Experiments \\
    27 & \cite{primary-study-54} & Shrinkage Estimators in Online Experiments \\
    28 & \cite{primary-study-55} & A Counterfactual Framework for Seller-Side A/B Testing on Marketplaces \\
    29 & \cite{primary-study-56} & Periodicity in User Engagement with a Search Engine and Its Application to Online Controlled Experiments \\
    30 & \cite{primary-study-60} & Evolving Software to be ML-Driven Utilizing Real-World A/B Testing: Experiences, Insights, Challenges \\
    31 & \cite{primary-study-67} & Using the Delay in a Treatment Effect to Improve Sensitivity and Preserve Directionality of Engagement Metrics in A/B Experiments \\
    32 & \cite{primary-study-68} & A Cluster-Based Nearest Neighbor Matching Algorithm for Enhanced A/A Validation in Online Experimentation \\
    33 & \cite{primary-study-77} & Variance-Weighted Estimators to Improve Sensitivity in Online Experiments \\
    34 & \cite{primary-study-80} & A Dirty Dozen: Twelve Common Metric Interpretation Pitfalls in Online Controlled Experiments \\
    35 & \cite{primary-study-82} & Winner's Curse: Bias Estimation for Total Effects of Features in Online Controlled Experiments \\
    36 & \cite{primary-study-84} & From Infrastructure to Culture: A/B Testing Challenges in Large Scale Social Networks \\
    37 & \cite{primary-study-89} & A Sequential Test for Selecting the Better Variant: Online A/B Testing, Adaptive Allocation, and Continuous Monitoring \\
    38 & \cite{primary-study-91} & Unbiased Experiments in Congested Networks \\
    39 & \cite{primary-study-94} & Personalized Treatment Selection Using Causal Heterogeneity \\
    40 & \cite{primary-study-100} & Beyond Success Rate: Utility as a Search Quality Metric for Online Experiments \\
    41 & \cite{primary-study-105} & Algorithms and System Architecture for Immediate Personalized News Recommendations \\
    42 & \cite{primary-study-115} & Experimentation in the Operating System: The Windows Experimentation Platform \\
    43 & \cite{primary-study-119} & AB4Web: An On-Line A/B Tester for Comparing User Interface Design Alternatives \\
    44 & \cite{primary-study-125} & Real-World Product Deployment of Adaptive Push Notification Scheduling on Smartphones \\
    45 & \cite{primary-study-139} & Mining the Stars: Learning Quality Ratings with User-Facing Explanations for Vacation Rentals \\
    46 & \cite{primary-study-143} & Towards Digital Twin-Enabled DevOps for CPS Providing Architecture-Based Service Adaptation \& Verification at Runtime \\
    47 & \cite{primary-study-145} & Adaptive Experimentation with Delayed Binary Feedback \\
    48 & \cite{primary-study-169} & Unifying Offline Causal Inference and Online Bandit Learning for Data Driven Decision \\
    49 & \cite{primary-study-174} & Beyond Data: From User Information to Business Value through Personalized Recommendations and Consumer Science \\
    50 & \cite{primary-study-189} & Learning to Bundle Proactively for On-Demand Meal Delivery \\
    51 & \cite{primary-study-196} & Measuring Dynamic Effects of Display Advertising in the Absence of User Tracking Information \\
    52 & \cite{primary-study-198} & Marketing Campaign Evaluation in Targeted Display Advertising \\
    53 & \cite{primary-study-209} & Whole Page Optimization: How Page Elements Interact with the Position Auction \\
    54 & \cite{primary-study-227} & The MOOClet Framework: Unifying Experimentation, Dynamic Improvement, and Personalization in Online Courses \\
    55 & \cite{primary-study-231} & Split-Treatment Analysis to Rank Heterogeneous Causal Effects for Prospective Interventions \\
    56 & \cite{primary-study-244} & Promoting Positive Post-Click Experience for In-Stream Yahoo Gemini Users \\
    57 & \cite{primary-study-245} & Predicting Counterfactuals from Large Historical Data and Small Randomized Trials \\
    58 & \cite{primary-study-277} & Multi-Source Pointer Network for Product Title Summarization \\
    59 & \cite{primary-study-278} & Beyond Relevance Ranking: A General Graph Matching Framework for Utility-Oriented Learning to Rank \\
    60 & \cite{primary-study-292} & Offline Evaluation to Make Decisions About PlaylistRecommendation Algorithms \\
    61 & \cite{primary-study-293} & Trustworthy Experimentation Under Telemetry Loss \\
    62 & \cite{primary-study-294} & The Netflix Recommender System: Algorithms, Business Value, and Innovation \\
    63 & \cite{primary-study-306} & Bifrost: Supporting Continuous Deployment with Automated Enactment of Multi-Phase Live Testing Strategies \\
    64 & \cite{primary-study-321} & CompactETA: A Fast Inference System for Travel Time Prediction \\
    65 & \cite{primary-study-327} & Design and Analysis of Benchmarking Experiments for Distributed Internet Services \\
    66 & \cite{primary-study-350} & Learning to Rank in the Position Based Model with Bandit Feedback \\
    67 & \cite{primary-study-351} & VisRel: Media Search at Scale \\
    68 & \cite{primary-study-355} & Behavioral Consequences of Reminder Emails on Students’ Academic Performance: A Real-World Deployment \\
    69 & \cite{primary-study-360} & Content Recommendation by Noise Contrastive Transfer Learning of Feature Representation \\
    70 & \cite{primary-study-362} & External Evaluation of Ranking Models under Extreme Position-Bias \\
    71 & \cite{primary-study-436} & Tackling Cannibalization Problems for Online Advertisement \\
    72 & \cite{primary-study-459} & Filling Context-Ad Vocabulary Gaps with Click Logs \\
    73 & \cite{primary-study-473} & Practical Lessons from Developing a Large-Scale Recommender System at Zalando \\
    74 & \cite{primary-study-476} & How Airbnb Tells You Will Enjoy Sunset Sailing in Barcelona? Recommendation in a Two-Sided Travel Marketplace \\
    75 & \cite{primary-study-482} & Modeling Professional Similarity by Mining Professional Career Trajectories \\
    76 & \cite{primary-study-506} & Social Incentive Optimization in Online Social Networks \\
    77 & \cite{primary-study-507} & Ad Close Mitigation for Improved User Experience in Native Advertisements \\
    78 & \cite{primary-study-561} & Off-Line vs. On-Line Evaluation of Recommender Systems in Small E-Commerce \\
    79 & \cite{primary-study-569} & LASER: A Scalable Response Prediction Platform for Online Advertising \\
    80 & \cite{primary-study-601} & The Role of Relevance in Sponsored Search \\
    81 & \cite{primary-study-614} & Contextual Bandit Applications in a Customer Support Bot \\
    82 & \cite{primary-study-615} & Safe Velocity: A Practical Guide to Software Deployment at Scale using Controlled Rollout \\
    83 & \cite{primary-study-636} & When Relevance is Not Enough: Promoting Diversity and Freshness in Personalized Question Recommendation \\
    84 & \cite{primary-study-668} & Interference, Bias, and Variance in Two-Sided Marketplace Experimentation: Guidance for Platforms \\
    85 & \cite{primary-study-676} & Automotive A/B testing: Challenges and Lessons Learned from Practice \\
    86 & \cite{primary-study-679} & A/B Testing at SweetIM: The Importance of Proper Statistical Analysis \\
    87 & \cite{primary-study-680} & A Framework Model to Support A/B Tests at the Class and Component Level \\
    88 & \cite{primary-study-681} & Statistical Reasoning of Zero-Inflated Right-Skewed User-Generated Big Data A/B Testing \\
    89 & \cite{primary-study-682} & Size matters? Or not: A/B testing with limited sample in automotive embedded software \\
    90 & \cite{primary-study-683} & Scalable Data Reporting Platform for A/B Tests \\
    91 & \cite{primary-study-693} & Applying Bayesian parameter estimation to A/B tests in e-business applications examining the impact of green marketing signals in sponsored search advertising \\
    92 & \cite{primary-study-694} & Experiment-driven improvements in Human-in-the-loop Machine Learning Annotation via significance-based A/B testing \\
    93 & \cite{primary-study-695} & The Anatomy of a Large-Scale Experimentation Platform \\
    94 & \cite{primary-study-697} & Demystifying dark matter for online experimentation \\
    95 & \cite{primary-study-699} & Controlled experiments for decision-making in e-Commerce search \\
    96 & \cite{primary-study-707} & Evaluating usability of a web application: A comparative analysis of open-source tools \\
    97 & \cite{primary-study-709} & Faster online experimentation by eliminating traditional A/A validation \\
    98 & \cite{primary-study-720} & Pitfalls of long-term online controlled experiments \\
    99 & \cite{primary-study-727} & SoftSKU: Optimizing Server Architectures for Microservice Diversity @Scale \\
    100 & \cite{primary-study-747} & The Benefits of Controlled Experimentation at Scale \\
    101 & \cite{primary-study-760} & Context Adaptation for Smart Recommender Systems \\
    102 & \cite{primary-study-763} & Whales, Dolphins, or Minnows? Towards the Player Clustering in Free Online Games Based on Purchasing Behavior via Data Mining Technique \\
    103 & \cite{primary-study-770} & Enterprise-Level Controlled Experiments at Scale: Challenges and Solutions \\
    104 & \cite{primary-study-785} & Should companies bid on their own brand in sponsored search? \\
    105 & \cite{primary-study-804} & A Probabilistic, Mechanism-Indepedent Outlier Detection Method for Online Experimentation \\
    106 & \cite{primary-study-835} & Inform Product Change through Experimentation with Data-Driven Behavioral Segmentation \\
    107 & \cite{primary-study-1159} & Your System Gets Better Every Day You Use It: Towards Automated Continuous Experimentation \\
    108 & \cite{primary-study-1312} & Fashion Recommendation Systems, Models and Methods: A Review \\
    109 & \cite{primary-study-1337} & Subject Line Personalization Techniques and Their Influence in the E-Mail Marketing Open Rate \\
    110 & \cite{primary-study-1453} & Impact of promotional social media content on click-through rate - Evidence from a FMCG company \\
    111 & \cite{primary-study-1476} & Related Entity Expansion and Ranking Using Knowledge Graph \\
    112 & \cite{primary-study-1495} & LinkLouvain: Link-Aware A/B Testing and Its Application on Online Marketing Campaign \\
    113 & \cite{primary-study-1572} & Ascend by Evolv: Artificial intelligence-based massively multivariate conversion rate optimization \\
    114 & \cite{primary-study-1587} & A new framework for online testing of heterogeneous treatment effect \\
    115 & \cite{primary-study-1623} & JACKPOT: Online experimentation of cloud microservices \\
    116 & \cite{primary-study-1636} & Digital Marketing Effectiveness Using Incrementality \\
    117 & \cite{primary-study-1679} & Business process improvement with the AB-BPM methodology \\
    118 & \cite{primary-study-1718} & A genetic algorithm for finding a small and diverse set of recent news stories on a given subject: How we generate aaai's ai-alert \\
    119 & \cite{primary-study-1733} & Measuring the value of recommendation links on product demand \\
    120 & \cite{primary-study-1796} & Experimentation growth: Evolving trustworthy A/B testing capabilities in online software companies \\
    121 & \cite{primary-study-1937} & Online Evaluation of Bid Prediction Models in a Large-Scale Computational Advertising Platform: Decision Making and Insights \\
    122 & \cite{primary-study-1956} & AB-BPM: Performance-driven instance routing for business process improvement \\
    123 & \cite{primary-study-1965} & Have It Both Ways---From A/B Testing to A{\&}B Testing with Exceptional Model Mining \\
    124 & \cite{primary-study-1966} & More for Less: Automated Experimentation in Software-Intensive Systems \\
    125 & \cite{primary-study-1970} & Regression Tree for Bandits Models in A/B Testing \\
    126 & \cite{primary-study-2026} & When the Crowd is Not Enough: Improving User Experience with Social Media through Automatic Quality Analysis \\
    127 & \cite{primary-study-2042} & Pixel efficiency analysis: A quantitative web analytics approach \\
    128 & \cite{primary-study-2055} & A/B Testing in E-commerce Sales Processes \\
    129 & \cite{primary-study-2060} & A Method for the Construction of User Targeting Knowledge for B2B Industry Website \\
    130 & \cite{primary-study-2066} & Validating Mobile Designs with Agile Testing in China: Based on Baidu Map for Mobile \\
    131 & \cite{primary-study-2088} & User Latent Preference Model for Better Downside Management in Recommender Systems \\
    132 & \cite{primary-study-2145} & Towards Automated A/B Testing \\
    133 & \cite{primary-study-2161} & Seven rules of thumb for web site experimenters \\
    134 & \cite{primary-study-2215} & Enabling A/B Testing of Native Mobile Applications by Remote User Interface Exchange \\
    135 & \cite{primary-study-2275} & Overlapping Experiment Infrastructure: More, Better, Faster Experimentation \\
    136 & \cite{primary-study-2287} & Optimizing price levels in e-commerce applications: An empirical study \\
    137 & \cite{primary-study-2293} & Facilitating Controlled Tests of Website Design Changes: A Systematic Approach \\
    138 & \cite{snowballed-study-1} & Soft Frequency Capping for Improved Ad Click Prediction in Yahoo Gemini Native \\
    139 & \cite{snowballed-study-2} & Objective Bayesian Two Sample Hypothesis Testing for Online Controlled Experiments \\
    140 & \cite{snowballed-study-3} & Test \& Roll: Profit-Maximizing A/B Tests \\
    141 & \cite{snowballed-study-4} & Improving Library User Experience with A/B Testing: Principles and Process
\label{tab:list-primary-studies}
\end{longtable}

\endgroup

\end{document}